\documentclass[aps,preprint,amsmath,amssymb,superscriptaddress]{revtex4}
\usepackage{epsf}
\usepackage{amsmath}
\usepackage{amssymb}
\usepackage{amsfonts}
\usepackage{appendix}
\usepackage{graphicx}
\usepackage{array}
\usepackage{setspace}
\usepackage{balance}
\usepackage{float}
\usepackage[bookmarks=true,hyperfigures=true]{hyperref}
\usepackage{siunitx}
\usepackage{caption}
\usepackage{subcaption}

\newcommand*{\brac}[1]{\left(#1\right)}

\DeclareMathOperator{\sgn}{sgn}

\begin{document}
    \title{Characterizing Memristive Nanowire Network Models via a Unified Computational Framework}
    
    \author{Marcus Kasdorf}
    \affiliation{Department of Physics and Astronomy, University of Calgary, 2500 University Drive NW, Calgary, Alberta T2N 1N4, Canada.}
    \affiliation{Institute for Quantum Science and Technology, University of Calgary, Calgary, Alberta, Canada.}
    \author{Diego Simpson-Ochoa}
    \affiliation{Department of Physics and Astronomy, University of Calgary, 2500 University Drive NW, Calgary, Alberta T2N 1N4, Canada.}
    \affiliation{Hotchkiss Brain Institute, University of Calgary, 3330 Hospital Drive NW, Calgary, Alberta T2N 4N1, Canada.}
    \author{Abdelrahman Bekhit}
    \affiliation{Schulich School of Engineering, University of Calgary, 2500 University Drive NW, Calgary, Alberta T2N 1N4, Canada.}
    \author{Mauro S. Ferreira}
    \affiliation{School of Physics, Trinity College Dublin, Dublin 2, Ireland.}
    \affiliation{Centre for Research on Adaptive Nanostructures and Nanodevices (CRANN) \& Advanced 
Materials and Bioengineering Research (AMBER) Centre, Trinity College Dublin, Dublin 2, Ireland.}
    \author{Wilten Nicola}
    \affiliation{Department of Physics and Astronomy, University of Calgary, 2500 University Drive NW, Calgary, Alberta T2N 1N4, Canada.}
     \affiliation{Hotchkiss Brain Institute, University of Calgary, 3330 Hospital Drive NW, Calgary, Alberta T2N 4N1, Canada.}
    \author{Claudia Gomes da Rocha}
    \affiliation{Department of Physics and Astronomy, University of Calgary, 2500 University Drive NW, Calgary, Alberta T2N 1N4, Canada.}
    \affiliation{Institute for Quantum Science and Technology, University of Calgary, Calgary, Alberta, Canada.}
    \affiliation{Hotchkiss Brain Institute, University of Calgary, 3330 Hospital Drive NW, Calgary, Alberta T2N 4N1, Canada.}
    
    \begin{abstract}        
        Randomly self-assembled nanowire networks (NWNs) are dynamical systems in which junctions between two nanowires can be modelled as memristive units viewed as adaptive resistors with memory. Various memristive models have been proposed to capture the complex mechanics of these junctions. Here, we showcase a novel computational framework named \texttt{Memristive Nanowire Network Simulator (MemNNetSim)} to simulate and analyze random memristive NWNs in a unified approach. Implemented using the Python programming language, \texttt{MemNNetSim} allows for the analysis of static and dynamic scenarios of NWNs under arbitrary memristive models. This provides a versatile foundation to build upon in further work, such as reservoir dynamics with NWNs, which has seen increased interest due to the interconnected architecture of NWNs. In this work, we introduce the package, demonstrate its utility in simulating NWNs, and then test advanced scenarios in which it can aid in the exploratory analysis of these systems, particularly in learning how to use NWNs as a physical reservoir in reservoir computing applications.
    \end{abstract}
    \maketitle
    
    \section{Introduction}

Conventional materials or synthetic matter with static physical properties can be viewed as primitive passive agents in various engineered and functional systems. However, such inert/canonical matter state has been supplemented by a new paradigm in materials science and nanoscience research with the ``rise of intelligent matter'' as coined by Kaspar et al. \cite{kaspar2021} and many other authors in the past decades \cite{maclennan2002,baulin2025,guo2023,grelz2023}. Intelligence is often associated with sentient beings as a result of their capacity to perceive, learn, and retain information, adapt and interact with the environment, and think or reason. With notable advances in artificial intelligence (AI) based technology, the notion of intelligence has extended beyond the realm of living organisms, with artificial computing systems being programmed to learn, adapt, and perform decision-making tasks akin to human reasoning. In an even more fundamental level, the raw materials that appear as building blocks in such computing architectures are also found to inherently exhibit a form of synthetic intelligence, meaning that cognition can be evidenced in synthetic matter or primitive hardware systems. In the context of intelligent matter or intelligent materials, these can exhibit smart/adaptable dynamical responses to external stimuli such as temperature, pressure, light, electric/magnetic fields, or chemical signals by self-regulating their internal properties accordingly, effectively learning from the environment and refining their function over time. A hallmark in the vast class of intelligent materials is the so-called memristive systems \cite{chua1971,chua1976}, matter characterized by adaptive resistive-switching responses rooted in the complex motion of elementary particle constituents such as atoms, ions, electrons, or spins that resemble the spiking of biological neuro-synapses. In particular, it has been shown that memristive systems can be used in in-memory computing strategies \cite{bao2022,lin2024} and to mimic synaptic plasticity \cite{ohno2011,sung2022,du2021,asif2023}, including spike-timing-dependent plasticity (STDP) \cite{serrano2013,bill2014,diederich2018,emelyanov2020,prezioso2016,serb2016,covi2016}, due to the possibility of systematic modulation of their dynamical conductance through pre- and post-synaptic-like voltage pulses. This sets their utilization in innovative learning rule designs equivalent to spiking neural networks \cite{saighi2015}. 

The foundational mathematical formulation for memristive systems was introduced by Chua \cite{chua1971} in 1971, who proposed the idealized ``memristor'', a lumped nonlinear resistive circuit element with memory driven by a single internal state variable, e.g., charge, following by its generalization to memristive systems with a multi-internal variable dynamics description by Chua and Kang in 1976 \cite{chua1976}. The seminal experimental realization pointing that the ``missing memristor'' was finally ``found'' came in 2008 with the work done in the Hewlett-Packard (HP) Labs \cite{strukov2008missing} that evidenced memristive signatures, characterized by current-voltage pinched hysteresis loops \cite{chua2012}, in TiO$_2$ nanostructures. Another prominent outcome of the work by Strukov et al. \cite{strukov2008missing} was the direct link between their experimental findings and a simplified memristor model they proposed to capture the complex charge-carrier transport and motion of ions (oxygen vacancies), the mobile agents responsible for the switching properties in the TiO$_x$-based devices. Since then, the nonlinearity, in-memory processing, adaptive, low-energy consumption, and synapse-like function evidenced in a wide variety of memristive nanomaterials and devices have confirmed the ample application of these intelligent solid-state systems in neuromorphic computing \cite{boybat2018,milano2018,kelly2016,wang2019,huang2020,sung2018,saighi2015,xiao2024}, in which hardware design is brain-inspired.

With its broad application range, memristive phenomena are known to be material-specific and are ruled by a wide variety of complex switching mechanisms and materials interface effects that can be grouped into four major groups \cite{waser2009,song2023,ventra2011,shao2025}: ionic, thermal, electronic, magnetic, and phase change. A typical memristive junction is composed of a metal-insulator-metal stack in which resistance modulation is achieved due to the formation/dissolution of a conductive filament (or boundary) inside the insulator in response to an external driving field. Such variability in the conductive filament state enables the memristive junction to switch discretely (binary) or continuously (analogue) between a high resistance state (HRS or OFF state) and a low resistance state (LRS or ON state). Some of the most studied memristive mechanisms at microscopic level are valence change memory (VCM) \cite{waser2021}, electrochemical metallization memory (ECM) \cite{yang2013}, phase-change memory (PCM) \cite{wright2011}, electronic (de)trapping-assisted processes \cite{guan2011} as well as quantum conductance mechanisms \cite{milano2022v3}, and spin-based switching \cite{shao2025}; each mechanism featuring its own mathematical/modelling framework, often phenomenologically devised. Beyond single memristive characterization, integration and collective behaviour of multiple memristive junctions have been achieved with state-of-the-art nanofabrication, with the onset of memristive arrays \cite{li2018,huang2020ann}, e.g., the crossbar layout \cite{xia2019} being significantly widespread, and random nanowire networks (NWNs) \cite{rocha2015ultimate,manning2018emergence,milano2025,loeffler2023}. In these systems, memristive effects emerge at the inter-wire connections from which bio-inspired learning can be harvested. Memristive arrays and random NWNs are structurally different, with the ``memarrays'' exhibiting ordered grid layouts, whereas random NWNs evidence disorder with the spaghetti-like assembly of the nanowires. They both exhibit pros and cons in terms of their use in brain-inspired applications or emulating artificial neural networks (ANNs) functions at the hardware level. A well-known disadvantage of crossbar arrays is the so-called sneak path (leakage) current effect \cite{shi2020,chen2021} that can impact read/write accuracy and operations, whereas random NWNs are immune to such leakages due to their inherent connectivity disorder, prone to fault-tolerant tasks. However, such disorder also brings challenges. For instance, one cannot precisely assess or extract information about the electrical state of single junctions, as the whole system is entangled in a complex interconnect of dynamical entities interacting with each other via electrical signals. In a way, random NWNs can be viewed as black boxes, difficult to infer or predict the internal state of the system that is also time-dependent. Yet, it is this rich combination of complexity, memory, and adaptive resistive behaviour that makes random NWNs attractive in prominent neuromorphic developments as summarized in the brief state-of-the-art subsection below.

\subsection{Neuromorphic Nanowire Networks State-of-the-art}
Earlier developments in hardware neuromorphics using random NWNs were presented by Stieg et al. \cite{stieg2012} and Avizienis et al. \cite{avizienis2012}, in which the authors identified remarkable collective and spatio-temporal brain-like responses of atomic switch networks made of interconnected Ag nanowires, pointing also to their potential for unconventional computing strategies. Theoretical and experimental demonstrations of the use of NWNs for reservoir computing were followed by Silling et al. \cite{sillin2013theoretical} and Demis et al. \cite{demis2016}, who showed the recurrence and high harmonic generation capability of random NWNs, acting as nonlinear reservoirs, to learn how to perform waveform transformations. Understanding the real-time transport dynamics of random NWNs and their internal mechanisms of distributing current through their network frame was theoretically and experimentally described by Manning et al. \cite{manning2018emergence}. In that work, we identified NWN self-similar behaviour and that under certain physical conditions, NWNs operate at the ``winner-takes-all'' (WTA) conduction state, characterized by the lowest power consumption connectivity state in the network. In other words, NWNs at a WTA state channel input electrical currents through highly selective pathways after their various memristive nanowire junctions evolve and ``learn'' that this is the most efficient (energetically) way of propagating current through their network skeleton. Subsequently, we also modelled and investigated their capacitive and transient transport regime, to reinforce the physical conditions which lead to the formation of WTA states in random NWNs in \cite{ocallaghan2018collective}. Multi-terminal current pathway control and visualization, plus memory/associative functionality in NWNs, were further presented by Li et al. \cite{li2020} and \cite{diaz-alvarez2020associative}, followed by the systematic demonstration of NWN structural plasticity due to reweighting and rewiring effects originating from their adaptable memristive junctions conducted by Milano et al. \cite{milano2020}. Information and network/graph theory were applied by Loeffler et al. \cite{loeffler2020}, Zhu et al. \cite{zhu2021}, Milano et al. \cite{milano2022}, and Montano et al. \cite{montano2022} to investigate dynamical information flow in NWNs, shedding light on how the topological structure of the network and hub sections play an important role in their smart information processing and memory capabilities. Adaptive complex behaviour in neuromorphic NWNs was observed by Scharnhorst et al. \cite{scharnhorst2018}, who evidenced features associated with self-organization, nonlinearity, and power law dynamics experimentally, corroborated by the study of Diaz-Alvarez et al. \cite{diaz-alvarez2019}, who further analyzed emergent behaviour as well as short-term memory and robustness of current pathway states in electrically activated NWNs. Two prominent review articles by Kunic et al. \cite{kunic2021} and Dunham et al. \cite{dunham2021}, released in 2021, offer valuable resources and a rich catalogue of developments in the use of random NWNs as neuromorphic entities, including their utilization as benchmark systems to study emergent phenomena equivalent to those observed in biological and complex systems. In particular, Michieletti et al. \cite{michieletti2025} further evidenced that memristive NWNs can exhibit self-organized and programmable local critical dynamics, which can be exploited in reservoir computing paradigms. Reservoir computing applied to waveform transformation tasks in NWNs was also studied by Daniels et al. \cite{daniels2022}, who also considered the often neglected nanowire (3D) stacking effects in their models. In another successful utilization of in-materia reservoir computing, Lilak et al. \cite{lilak2021} built a successful neuromorphic NWN reservoir for spoken digit classification. Further prominent studies exploiting the emergent, nonlinear dynamics and self-organization characteristics in random NWNs are highlighted here. Milano et al. \cite{milano2022v2} fully realized {\it in materia} reservoir computing \cite{fang2023}, and subsequently incorporated stochastic effects into their memristive modelling framework \cite{milano2025}, from which noise and conductance fluctuations were successfully derived with excellent agreement with their experimental findings. Lastly, two major breakthroughs in the demonstration of the synthetic cognitive features of NWNs came with the works of Loeffler et al. \cite{loeffler2023} and Zhu et al. \cite{zhu2023}. In these works, the authors successfully implemented supervised and reinforcement learning tests in NWNs, as well as $n$-back recognition tasks to probe their capacity for sequential memorization \cite{loeffler2023}, and accurate online learning to conduct image classification tasks applied to the MNIST handwritten dataset \cite{zhu2023}.

Inspired by all these developments in neuromorphic and reservoir computing applied in random NWNs, in this work, we devised a first version of a unified and flexible computational platform \cite{kasdorf_github}, capable of modelling and testing a number of {\it in-materia} memristive behaviour using NWNs as intelligent reservoirs. From the comprehensive literature review above, the potential for neuromorphic hardware using disordered NWNs is certainly identified, but challenges exist regarding standardization and benchmarking, scalability, and lack of well-established protocols to control their electrical responses to guarantee reproducible outcomes. Robust systematic studying to account for the immense variety of brain-inspired tests, as well as the large amount of possible configurational arrangements and circuit design, materials choice, diverse memristive models and mechanisms, and enhanced parametrical phase-space to span whenever NWNs are studied in this context, demands a flexible and open-source simulation platform to enable direct changes in the modelling framework. In this way, we introduce the \texttt{Memristive Nanowire Network Simulator (MemNNetSim): A neuromorphic proof-of-concept modular toolkit} \cite{kasdorf_github,kasdorf_github_io,kasdorf_pypi} from which users can set the NWN initial materials and spatial conditions, test three distinct memristive models already implemented in the package to emulate the interwire dynamical junctions, and simulate their adaptive electrical response in time. The toolkit is written in Python programming language \cite{python} with some snippets accessible in Jupyter Notebook with documentation, comprehensive user-guide, and demonstration of our results probing major ingredients essential for neuromorphic computing, inspired by the cited literature above: evidence of current-voltage hysteresis loops, power spectral analysis of current potentiation to identify power law behaviour, bit pattern storage through activation of selective pathways, associative memory through multi-electrode stimulation, and reservoir-based waveform transformation. Through such a transparent computational and mathematical toolkit, we can predict and explain a wide variety of novel intelligent behaviour in random NWNs and how they couple with materials' intrinsic and extrinsic properties.

In the following, we detail the main memristive models adopted to describe the switching properties at the interwire junctions, as well as their integration into the NWN circuit framework. We also briefly discuss the computational package workflow and its modular structure. In the Results section, we showcase a wide range of simulations that \texttt{MemNNetSim} can carry out, ranging from emerging spatiotemporal electrical dynamics to memory capabilities, finalizing with applicable waveform transformation based on memristive reservoir computing strategy. We wrap up our main findings and conclusions in the Conclusions section.

    \section{Methods}

\texttt{MemNNetSim} is written in the \texttt{Python} programming language \cite{python} with some demos and benchmark results written as Jupyter Notebooks available on the documentation website \cite{kasdorf_github,kasdorf_github_io,kasdorf_pypi}. Its main routines and modules are written following a hybrid of procedural and object-oriented programming (OOP) styles, from which NWN-related objects and simulations can be easily set to run, also with options for direct plotting visualization and statistical ensemble analysis. The main \texttt{Python} routines the code imports are \texttt{Matplotlib} for graphical results, \texttt{NetworkX} and \texttt{Shapely} for network and geometrical characterization of NWNs, \texttt{Scipy} and \texttt{NumPy} for various numerical routines and solvers, including ordinary differential equations (ODE) and generalized circuit solvers following Modified Nodal Analysis (MNA) \cite{ho1975modified,mna}. Starting from the NWN generation, nanowires are represented as 1D line segments with a fixed diameter on a 2D plane. To generate a nanowire of a given length, a coordinate for the midpoint and an angle to rotate the nanowire around its midpoint are randomly chosen. To generate an entire NWN, the user provides the effective device dimensions, i.e., the length and width of the region to generate nanowires in, along with a desired wire density defined as the number of wires per $\mu$m$^2$. Nanowires are continually randomly generated with midpoints within the given region until the wire density is attained. Once all nanowires have been created, the locations of the nanowires and all nanowire intersections are recorded.

\begin{figure}[H]
    \centering
    \includegraphics[width=1.0\linewidth]{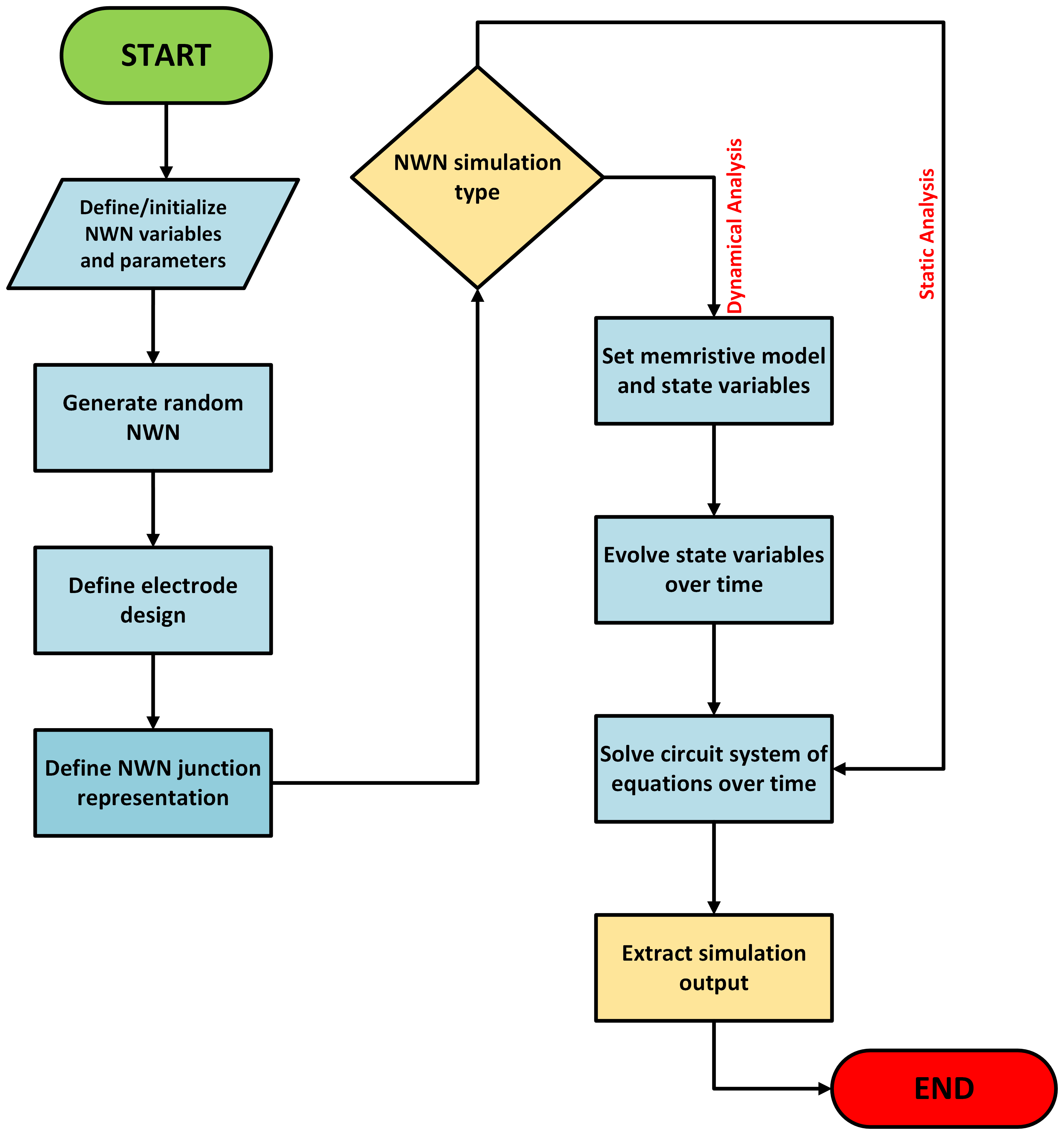}
    \caption{Workflow of \texttt{MemNNetSim} package detailing its main initialization procedures and simulation types: dynamical/time-dependent or static transport analysis in memristive random NWNs \cite{kasdorf_github,kasdorf_github_io,kasdorf_pypi}.}
    \label{fig:workflow}
\end{figure}

Figure \ref{fig:workflow} presents the code workflow of the package, consisting of initialization procedures in which the user can define NWN geometrical characteristics, e.g., wire length and diameter, wire density, materials properties, definition of units, and the choice of electrode design from which the network can be electrically interrogated by two or multiple terminals of voltage or current sources arranged around the device area. The whole NWN device is therefore generated through a pseudo-random number generator scheme that spreads the location of the wires and their orientations randomly over a pre-defined circuit area. Graph representation is used to efficiently store the relational data between the nanowires and their connections. The connectivity and nodal mapping of the NWN can be transcribed into two main graph representations introduced in our previous work at \cite{rocha2015ultimate}: the junction-dominated assumption (JDA) and the multi-nodal representation (MNR). JDA considers that the resistances at the junctions are much larger than the wires' internal resistances. As a result, nanowire junctions (or edges of the graph) are assumed to be the only source of resistance, with the nanowires themselves becoming an equipotential node. The resistance associated with the junctions is naturally stored as an edge attribute in the graph. Conversely, in a MNR NWN representation, nanowires do have internal resistances. For this type, graph nodes represent contact points on a wire, and the graph edges represent either nanowire intersections or nanowire segments. In this package, the default NWN representation is JDA, with the option to convert to MNR.

Once all the nanowire, electrode, and nanowire junction information have been stored in the graph, MNA \cite{ho1975modified,mna,kahale2022memristor} can be performed to obtain the voltage at every node in the graph. This has the advantage over standard nodal analysis, as the MNA matrix equation for a general network of circuit elements can be directly obtained from the NWN graph. Furthermore, we can freely choose nodes to be an ideal source or ground without a lack of generalization. Once source and ground nodes are chosen, the static configuration of the NWN is completed, and a static analysis of the NWN can be performed. Via MNA, all the nodal voltages are calculated and returned. This static analysis is the foundation for the dynamic analysis, in which the key difference is how the junction resistances are modelled. By default, the junctions are assumed to be static and implemented with standard resistors. This assumption is lifted in the dynamic case, where junctions are instead modelled as memristive units operating as adaptive resistors with memory, with their initial conditions being set in the first iteration of the static case simulation. Memristive elements are able to adapt their resistance based on a number of internal state variables that change depending on their present state and current flowing through them \cite{chua1971,chua1976}. They are governed by a modified version of Ohm's law in which the resistance is a function of these variables:
\begin{align}
    v &= R(\vec x, i) i
\end{align}
where $v$ is the voltage, $R$ is the resistance, $i$ is the current, and $\vec x$ is a vector of state variables. The (dynamical) state variables can be modelled over time, $t$, following a rate equation of the form
\begin{align}
    \frac{d\vec x}{dt} = \vec f(\vec x, i) \, 
\end{align}
with $\vec{f}$ being functions that govern how the state variables change with time. In the \texttt{MemNNetSim} package, both the resistance function and the state equations need to be provided to simulate the dynamics of a NWN. This is where the package offers flexibility, in which the user can choose which memristive model to use, following a modular scheme with freedom to add customized response and differential equations. A restriction when providing these functions is that they must be \textit{dimensionless}, that is, they have the units removed. All calculations are performed in arbitrary units so that the same algorithms can be used regardless of their units. This also has the notable advantage that simulation results are able to be scaled to any desired size since the calculations for the evolution and MNA are standardized across all models.

\texttt{MemNNetSim} comes with three benchmark memristive models pre-implemented: the linear drift model introduced by Strukov et al. \cite{strukov2008missing}, referred here as the ``HP model'', a version of the linear drift model plus a decay/diffusion term \cite{chang2011,chen2013} as introduced by Sillin et al. \cite{sillin2013theoretical}, referred here as the ``Decay HP model'', and a second version of the linear drift model incorporating short- and long-term memory effects introduced by Chen et al. \cite{chen2014phenomenological}, referred here as the ``SLT HP model''. Starting with the HP model, the dimensionless resistance and state equation that describe it are:
\begin{gather}
    v(t) = R[x(t)]\, i(t) \,\,\,\mbox{with} \,\,\, R(x) = x\brac{1 - \frac{R_\text{off}}{R_\text{on}}} + \frac{R_\text{off}}{R_\text{on}} 
    \label{eq:linear-response} \\
    \frac{dx}{dt} = i(t)h(x)
    \label{eq:HP-model}
\end{gather}
where $x$ is the internal state variable, $R_\text{on}$ is the resistance of the ON state and $R_\text{off}$ is the resistance of the OFF state. It should be noted that $R_\text{on}$ here is the characteristic unit for the resistance, and so the ratio of $R_\text{off}/R_\text{on}$ is unitless. This expression is only valid when the state variable $0\leq x \leq 1$. In the original HP model \cite{strukov2008missing}, for instance, $x=w/d$ in which $w$ is the length of the doped boundary and $d$ is the total length of the memristive channel. A window function $h(x)$ is also often added so that the boundary behaviour of $x$ is well-defined and correctly constrained. Strukov et al. \cite{strukov2008missing} addressed this point with their window function
\begin{align}
    h(x) = \begin{cases}
        x (1 - x) & \text{for } 0 \leq x \leq 1 \\
        0 & \text{otherwise}
    \end{cases}
    \label{eq:strukov-window}
\end{align}
which added a nonlinear drift to $x$ when it approaches either zero or one. For more control over the speed at which the state equation is suppressed at the boundaries, Joglekar \& Wolf \cite{joglekar2009} introduced a family of window functions parameterized by a positive integer $p$ of the form:
\begin{align}
    h_p(x) = \begin{cases}
        1 - (2x - 1)^{2p} & \text{for } 0 \leq x \leq 1 \\
        0 & \text{otherwise}
    \end{cases}
    \label{eq:joglekar-window}
\end{align}
When using the \texttt{MemNNetSim} package, the window function is a required parameter for the pre-implemented models.

Building on the HP model as a basis, the Decay HP model adds a decay term to the state equation of $x$ to incorporate the aspect of a memristor ``forgetting'' over time \cite{chang2011,chen2013,zhou2019}. The strength of this decay is characterized by the decaying rate constant $\tau$, and the governing state equation becomes
\begin{align}
    \frac{dx}{dt} = i(t)h(x) - \frac{x}{\tau} \, .
    \label{eq:decay-model}
\end{align}
The response function used for the Decay model is the same function as for the HP model, as in Equation \eqref{eq:linear-response}. 

The SLT HP model further builds on the Decay model and adds a state equation for $\tau$ to allow it to change over time, changing it from a constant to a state variable. It also incorporates the aspect of a memristor, reinforcing its memory of previous states with the state variable $\varepsilon$, also referred to as retention, to account for long-term memory. This leads to the three-state equations for the model as proposed by Chen et al. \cite{chen2014phenomenological},
\begin{equation}
    \begin{aligned}
        \frac{dx}{dt} &= \brac{i(t) - \frac{x - \varepsilon}{\tau}}h(x) \\
        \frac{d\tau}{dt} &= i(t) \theta (a - x) \\
        \frac{d\varepsilon}{dt} &= i(t) \sigma h(x) \, .
    \end{aligned}
    \label{eq:chen-model}
\end{equation}
It is seen that the strength of the current flow impacts $\tau$ and $\varepsilon$, governed by the model parameters $\theta$ and $\sigma$, respectively. There is an additional model parameter $a$ that also affects whether the current flow is positively or negatively correlated to $\tau$.

\begin{figure}[H]
    \centering
    \includegraphics[width=0.85\linewidth]{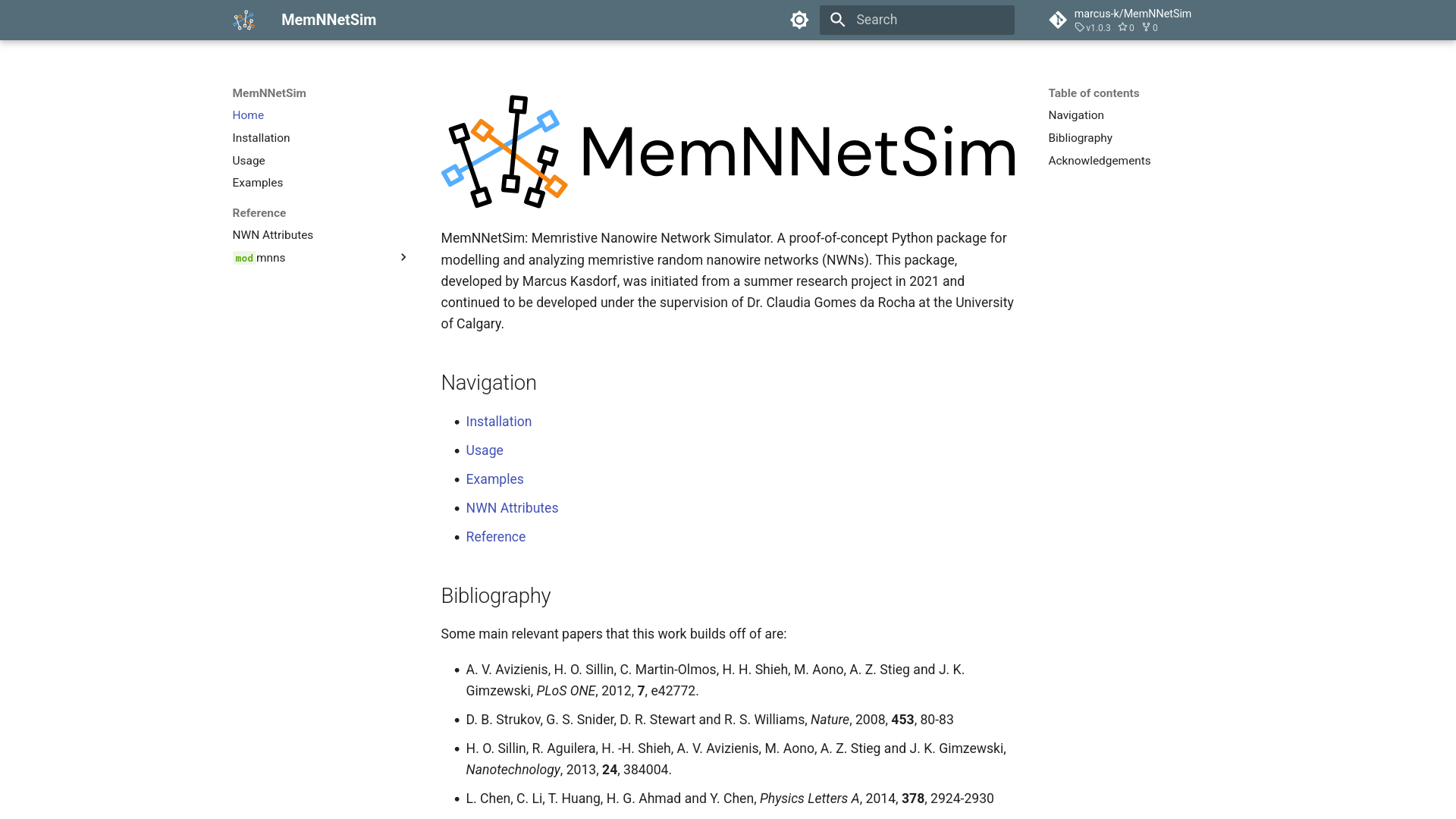}
    \caption{Snapshot of the webpage hosting the complete \texttt{MemNNetSim} package in GitHub \cite{kasdorf_github_io}. The package and documentation can also be accessed here \cite{kasdorf_github,kasdorf_pypi}. These pages and the software are an ongoing work and will continue to be updated to different versions, extensions, and other model considerations.}
    \label{fig:webpage}
\end{figure}

\texttt{MemNNetSim} is not limited to these three models, however. One has the ability to provide custom resistance response functions and state equations for full control over the dynamic simulations. As these equations are only the memristive models for a single memristive junction between nanowires, the package applies the desired model across an entire network of randomly distributed nanowires. The memristive equations govern the time evolution of the state and response functions of the junctions in the NWN initially set, and those are paired with MNA to get the complete circuit dynamics that can be analyzed and plotted from the package output data. Emergent and nonlinear dynamics will be analyzed in the following Results section, composed as a catalogue of a series of neuromorphic-like computational experiments we designed and included in the package tutorials \cite{kasdorf_github,kasdorf_github_io,kasdorf_pypi}. A snapshot of the webpage hosting the complete package (linked with GitHub) is shown in Figure \ref{fig:webpage}, together with many other resources, information, tutorials, main references, and tests. The webpage is created with \texttt{MkDocs} \cite{mkdocs}, a themeable \cite{mkdocs-material} static site generator with the added ability to automatically generate documentation pages from Python docstrings via \texttt{mkdocstrings} \cite{mkdocstrings}. This page is an ongoing work and will continue to be updated as extensions, further tests, and other features are added to the package.

    \section{Results and Discussion}

Before probing dynamical responses and memristive capabilities in random NWNs using \texttt{MemNNetSim}, we introduce in the supplementary information \cite{supplementary} a set of benchmark static transport -- sheet resistance ($R_s$) predictions -- conducted in systems of various wire densities following an ensemble analysis. The static analysis can provide an initial inquiry into the materials' and network geometrical settings that can be used as initial conditions for the dynamical analysis. NWN device dimensions, nodal representation, inner ($R_{in}$) and junction ($R_j$) resistances, wire density ($n_w$), as well as nanowire diameter ($D$) and length ($L$) are set based on experimental characterizations conducted in our past works \cite{rocha2015ultimate,bellew2015resistance,ocallaghan2018collective}. A summary of the parameters set and carried on to most of the results presented here are: nanowire resistivity of $\SI{22.6}{n\ohm m}$ for Ag, $D=50$ nm, $L=7$ $\mu m$, wire densities typically $n_w>0.1$ nanowires$/\mu m^2$ to guarantee percolation \cite{li2009finite}, device sizes of at least $20\times 20$ $\mu m$, and for meaningful statistical results, NWN ensembles contain at least 5000, up to 10000 samples.

\begin{figure}[H]
    \centering
    \includegraphics[width=\linewidth]{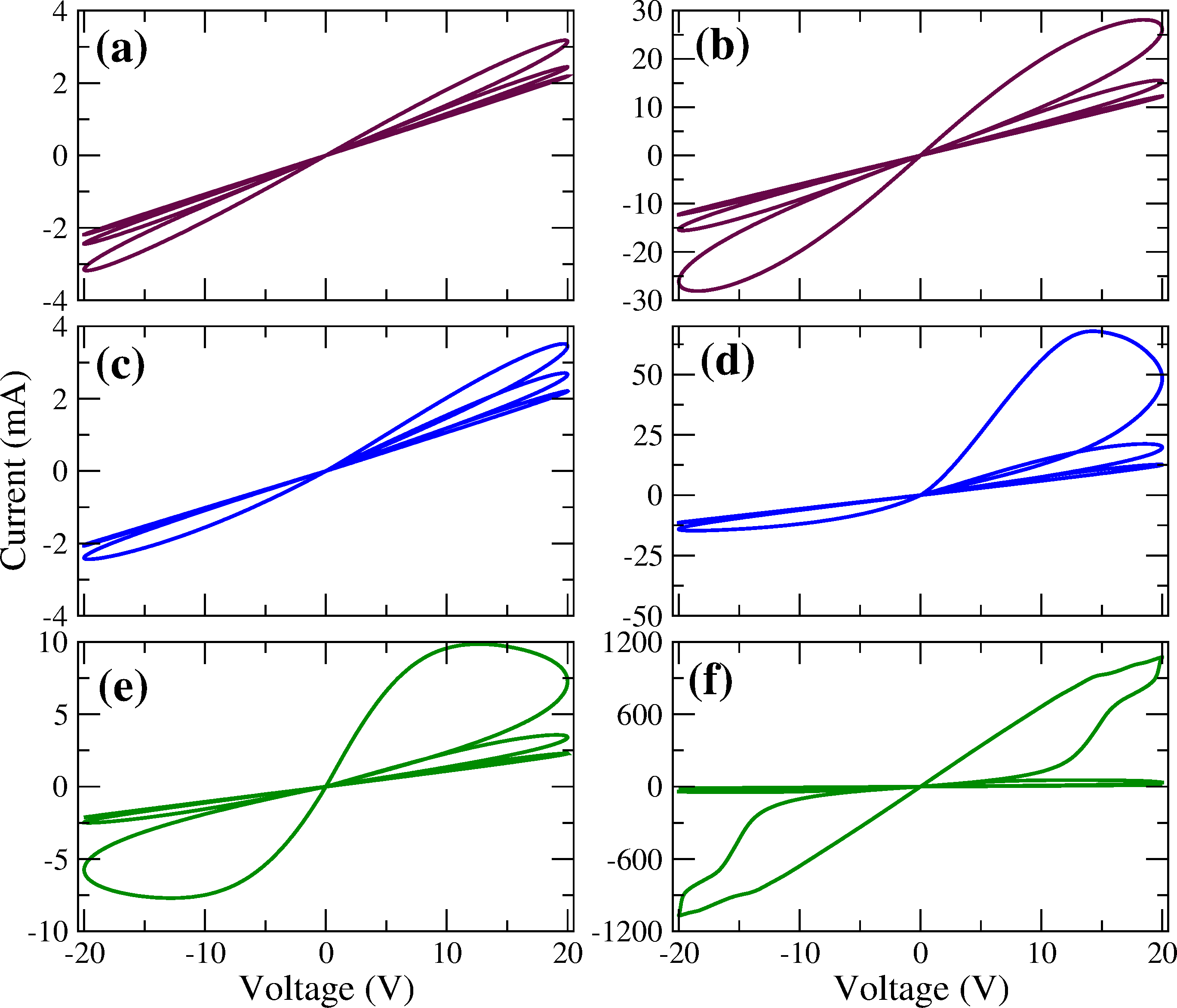}
    \caption{Voltage versus current (pinched) hysteresis loops for a random NWN with two wire density values and distinct memristive models. Results for a NWN of wire density 0.12 $\mu$m$^{-2}$ (sparser case) are depicted on the left panels, whereas for a NWN of wire density 0.20 $\mu$m$^{-2}$ (denser case) are depicted on the right panels. The hysteresis shown on the top (a,b) panels are for the HP model (simple linear drift), the mid panels (c,d) are for the Decay HP model, and the bottom (e,f) panels are for the SLT HP model. Three positive and negative voltage (half-cycles) sweeps within the $[-20,20]$ V voltage range are displayed, with the inner area of the hysteresis increasing with time evolution, evidencing that the baseline conductance of the NWN has improved upon repeated voltage cycling. All loops are counter-clockwise (CCW) in the positive voltage range and clockwise (CW) in the negative voltage range.}
    \label{fig:hysteresis-comparison}
\end{figure}

\autoref{fig:R_s-vs-R_j} depicts the sheet resistance dependence on the junction resistance value set for all junctions in a NWN of a given density, made of Ag nanowires, following the JDA and MNR nodal representations for comparison. It is expected a linear dependence in both representations, with the linear fit indicating a pronounced intercept parameter for the case of MNR to reflect the influence of the nanowire inner resistances. This contrasts with JDA exhibiting a nearly zero intercept since this representation does not consider the influence of the nanowire inner resistances. Another expected static transport result from the package is the scaling law of $R_s$ with wire (critical) density. This is shown and further discussed in the supplementary information \autoref{fig:R_s-vs-n_w} from which the characteristic scaling relation $R_s \propto (n_w - (n_w)_c)^{-\alpha}$ can be observed, with fitted exponent $\alpha\approx 1.2$ and critical wire density $(n_w)_c\approx 0.11$ nanowires$/\mu m^2$ matching studies conducted in stick percolation systems \cite{li2009finite,li2010conductivity,sahimi1983critical,balberg1983critical}. These results serve as ``sanity checks'' to demonstrate that all integrated modules, libraries and routines are functioning well and predicting the expected sheet resistance trends under static transport regime. Further details and analysis on Figures S1 and S2 are provided in the supplementary information \cite{supplementary}.

The interesting nonlinear emergent neuromorphic utilities of memristive random NWNs are evidenced when dynamical simulations are enabled in \texttt{MemNNetSim}. The main memristive blueprints of the systems, i.e., current versus voltage pinched hysteresis loops, are depicted in Figure \ref{fig:hysteresis-comparison} for distinct nanowire densities simulated using each of the memristive models introduced in the previous section: the HP model, the Decay HP model, and the SLT HP model. A series of positive and negative current-voltage (half-cycles) sweeps, including the quasi-steady-state, is presented in the figure from which we can identify important transport signatures that will be relevant in interpreting results and applications of the NWNs as reservoirs later on. A first important outcome from this result is the impact of the nanowire density on the shape of the hysteresis loops. The sparser NWN exhibits simpler loops like an ``ideal memristor'', whereas the denser NWN depicts an asymmetric pinched hysteresis with the Decay HP model and a butterfly-like shape with the SLT HP model after a number of cycles. Differences and signatures of each model are therefore better perceived in denser NWNs.

\begin{figure}[H]
    \centering
    \includegraphics[width=0.9\linewidth]{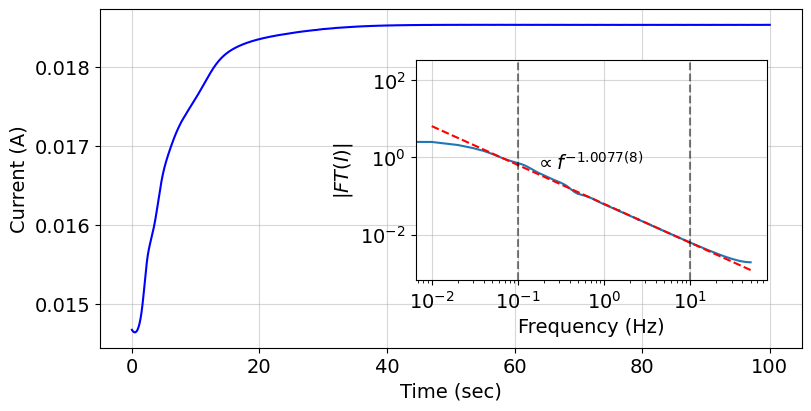}
    \caption{(Main panel) Current evolution of a JDA NWN using the Decay HP model stressed with \SI{20}{V} DC for 100 seconds. The NWN dimensions are $\SI{56}{\micro m} \times \SI{35}{\micro m}$, all nanowire lengths were set at \SI{7}{\micro m}, and a nanowire density of \SI{0.3}{\micro m^{-2}} was maintained. A four-terminal electrode configuration was used, with two electrodes placed on the left side of the NWN and two others on its right side. The characteristic resistance of $R_\text{on} = \SI{10}{\ohm}$ and resistance ratio of $R_\text{off}/R_\text{on} = 160$ were used in this result. (Inset) Log-Log plot of the absolute value of the Fourier transform of the main panel response. The dashed red line is the power law fitting of a function $y \propto f^{-\beta}$ conducted within the frequency ($f$) range delimited by the vertical dashed lines. The fitting gives an exponent of $\beta = \SI{1.0077\pm0.0008}{}$.}
    \label{fig:I-vs-t}
\end{figure}

\begin{figure}[H]
    \centering
    \includegraphics[width=0.7\linewidth]{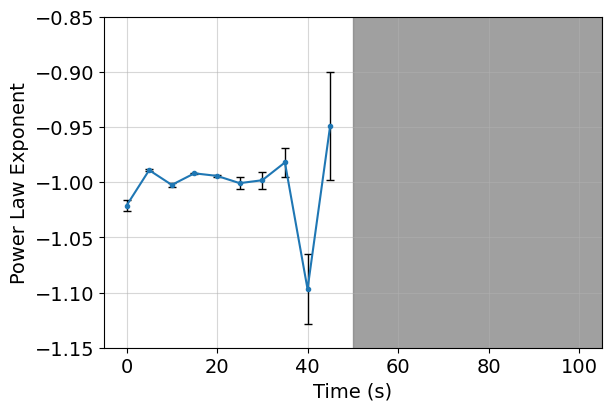}
    \caption{Power law exponent $\beta$ fitted from the Fourier transformed current response in \autoref{fig:I-vs-t} for a NWN modelled with the Decay HP model, partitioned into time windows of 5 seconds. The grey section indicates the region in which the fitted exponent had an uncertainty larger than 0.1, and therefore its values are not representative of a power law.}
    \label{fig:beta-vs-t}
\end{figure}

\begin{figure}[H]
    \centering
    \includegraphics[width=\linewidth]{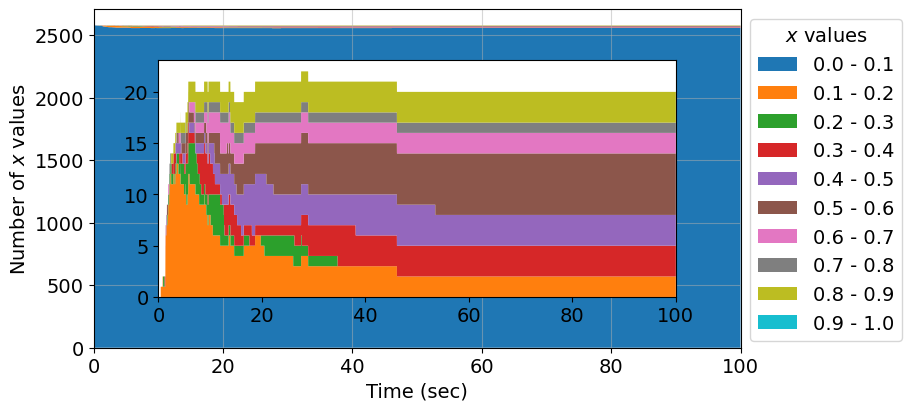}
    \caption{State variable $x$ for all memristive NWN junctions, recorded during the current simulation in \autoref{fig:I-vs-t} for a NWN modelled with the Decay HP model, binned into intervals of 0.1 as a function of time. The inset (central panel) shows the same binned data with the 0.0 - 0.1 bin omitted for improved visualization.}
    \label{fig:stacked-w-decay}
\end{figure}

Another important dynamical response to simulate in memristive NWNs is the current evolution at constant bias voltage stimulus. This result allows us to track the NWN activation and its current adaptability over time. First, we investigated the current time evolution of NWNs governed by the Decay HP model for the nanowire junctions. The NWN used here had $\SI{56}{\micro m} \times \SI{35}{\micro m}$ in device dimensions, JDA nodal representation, all nanowire lengths set at \SI{7}{\micro m}, and a nanowire density of \SI{0.3}{\micro m^{-2}}. The random number NWN generator had a fixed seed of 5 for reproducibility. Four electrodes were arranged on the lateral edges of the NWN area; two electrodes were placed on the left side, and two others were placed on the right side. The same characteristic resistance of $R_\text{on} = \SI{10}{\ohm}$ and resistance ratio of $R_\text{off}/R_\text{on} = 160$ were used for all nanowire junctions for scaling the results. A DC voltage of $\SI{20}{V}$ was applied between the bottom-left and top-right electrodes, and Strukov's \cite{strukov2008missing} window function was used in the state dynamical equations. A decay constant of $\tau = \SI{1}{s}$ was chosen, and all the nanowire junctions of the NWN were initialized at $x = 0.05$. The NWN was exposed to the constant bias voltage or electrically stressed for \SI{100}{s}, with the resulting current flow across it over time graphed in \autoref{fig:I-vs-t}. This duration was chosen to guarantee enough electrical exposure until we observe the current plateauing as seen in the figure. The inset figure shows the Fourier transformed current response across all \SI{100}{s}. Fitting the power spectrum data within the frequency range of \SI{1e-1}{Hz} and \SI{1e1}{Hz}, a power law trend was realized following a $f^{-\beta}$ scaling with $f$ being the frequency and $\beta = \SI{1.0077\pm0.0008}{}$. This power law trend, however, is only valid within these two decades of frequency due to finite size effects related to the fact that the current response exhibits a lower cutoff and plateaus after a certain duration of being stressed. Yet, we can observe a clear scale-free behaviour within two frequency decades, and these simulations are in agreement with the experimental realizations, as we will discuss later. The persistence of this power law was further analyzed in \autoref{fig:beta-vs-t}, where the current response was broken up into 5-second intervals and in each interval the critical exponent $\beta$ was fitted to the same two-decade range of frequencies. An uncertainty limit for $\beta$ of $0.1$ was set, which at the point $\beta$ was said to no longer accurately represent the data. For this simulation, this cutoff occurred at the $\SI{50}{s}$ mark. This was a similar time when the current showed plateauing in \autoref{fig:I-vs-t}. To analyze this connection further, \autoref{fig:stacked-w-decay} illustrates how the state variable $x$ of each nanowire junction changed as a function of time in a binned colored diagram. Here, the cutoff corresponds to the point where the $x$ values achieve a steady state and are no longer actively changing. From these results, stressing and training the NWN appears to place the NWN in a state of criticality for the duration of training, after which the power law no longer holds and the NWN achieves a steady state with respect to the applied voltage. The term ``training'' here refers to the NWN learning the task of propagating current in an energy-efficient manner by dynamically adjusting or adapting the resistance state of its memristive junctions so that selected paths of least resistance can be activated. Knowing that dissipated power is $P=V_{DC}^2/R_{eq}$ with $V_{DC}$ being the DC voltage applied across the NWN and $R_{eq}$ its equivalent resistance, the NWN will ``administer'' current flow following a WTA conduction strategy as we demonstrated in \cite{manning2018emergence} to reduce dissipated power. In such a state, the NWN does not rely on redundant (multiple) pathways to propagate current; it rather gradually activates a reduced number of hub junctions that facilitate the formation of least-resistive paths. Note also the reduced number of junctions with $x>0.8$, indicating that junctions near full ON activation state are rare events and that the dynamics in the state variable $x$ cease for $t>50$ s.

\begin{figure}[H]
    \centering
    \includegraphics[width=0.9\linewidth]{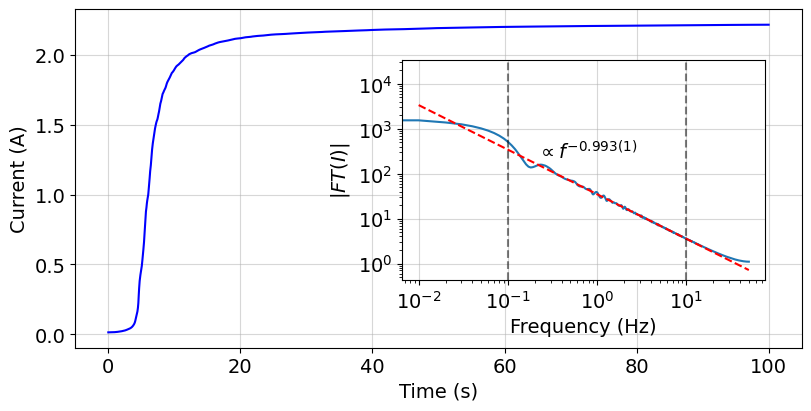}
    \caption{(Main panel) Current evolution of a JDA NWN using the SLT HP model stressed with \SI{20}{V} DC for 100 seconds. The NWN dimensions are $\SI{56}{\micro m} \times \SI{35}{\micro m}$, all nanowire lengths were set at \SI{7}{\micro m}, and a nanowire density of \SI{0.3}{\micro m^{-2}} was maintained. A four-terminal electrode configuration was used, with two electrodes placed on the left side of the NWN and two others on its right side. The characteristic resistance of $R_\text{on} = \SI{10}{\ohm}$ and resistance ratio of $R_\text{off}/R_\text{on} = 160$ were used in this result. (Inset) Log-Log plot of the absolute value of the Fourier transform of the main panel response. The dashed red line is the power law fitting of a function $y \propto f^{-\beta}$ conducted within the frequency ($f$) range delimited by the vertical dashed lines. The fitting gives an exponent of $\beta = \SI{0.993\pm0.001}{}$.}
    \label{fig:I-vs-t-chen}
\end{figure}

\begin{figure}[H]
    \centering
    \includegraphics[width=0.7\linewidth]{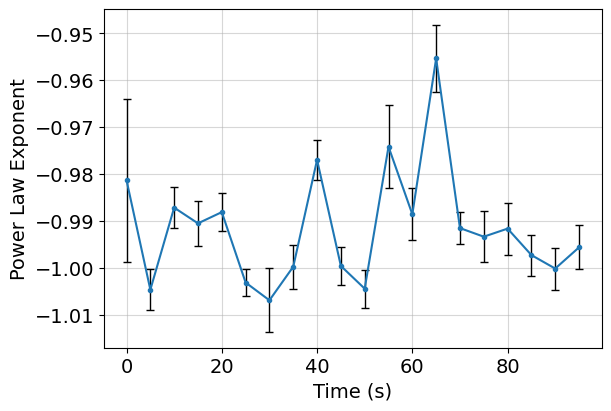}
    \caption{Power law exponent $\beta$ fitted from the Fourier transformed current response in \autoref{fig:I-vs-t-chen} for a NWN modelled with the SLT HP model, partitioned into time windows of 5 seconds. Power law persisted throughout the whole time period of the simulation, with meaningful fitted exponents.}
    \label{fig:beta-vs-t-chen}
\end{figure}

\begin{figure}[H]
    \centering
    \includegraphics[width=\linewidth]{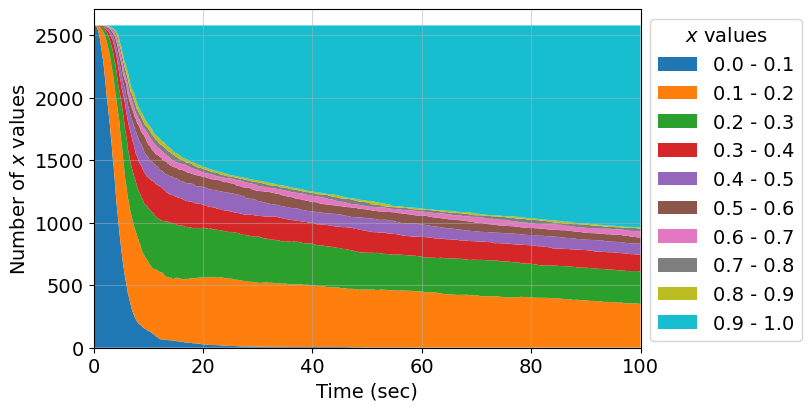}
    \caption{State variable $x$ for all memristive NWN junctions, recorded during the current simulation in \autoref{fig:I-vs-t-chen} for a NWN modelled with the SLT HP model, binned into intervals of 0.1 as a function of time.}
    \label{fig:stacked-w-chen}
\end{figure}

For comparison, the NWN was reset and electrically stressed again at DC 20 V using the SLT HP model. The initial state variable values for each nanowire junction were $x(t=0) = 0.05$, $\tau(t=0) = \SI{1}{s}$, and $\varepsilon(t=0) = 0.1$. Other parameters used for this model were $\sigma = 0.1$, $\theta = 0.1$, and $a = 0.9$ and the same voltage, characteristic ON resistance, and resistance ratio as the prior Decay HP mode simulation were applied. The current response is graphed in \autoref{fig:I-vs-t-chen}. The scaled current for this case is noticeably larger than the Decay HP model, highlighting the effect caused by the retention term in the model, coupled with a dynamical decaying rate. The inset displays the Fourier transform of the current response (main panel) from which a power law is observed covering two frequency decades, as in the Decay HP model, however, a small fluctuation in the fitted exponent $\beta$ was seen, with $\beta = \SI{0.993\pm0.001}{}$. Within a certain significance level, one can say that the $\beta$ values for the two models are comparable, nonetheless, at uncertainties smaller than 0.001, this difference reflects on a frequency power spectrum at the sublinear regime for the SLT HP model. This means the frequency power spectrum of the NWN with the SLT HP model decays more slowly than the NWN with the Decay HP model, and it is skewed towards higher frequencies. In other words, higher (lower) frequency modes are more (less) frequent in the NWN driven by the SLT HP than the Decay HP model. A power law trend was still observed in the Fourier-transformed current response in \autoref{fig:I-vs-t-chen} across the 100 seconds of applied DC voltage. Unlike the Decay HP model, \autoref{fig:beta-vs-t-chen} shows how the power law relationship persists across the entire training period of the NWN modelled with SLT HP. A close-up at the junctions' state variable binned over time is shown in \autoref{fig:stacked-w-chen}, revealing that a significant proportion of the nanowire junctions are fully or nearly activated ($x \in [0.9,1]$). This suggests that memristive junctions within the SLT HP model adapt quickly to the higher current flow, with various parallel trained paths composed of many junctions at the ON resistance state. On the other hand, this result also shows many nanowire junctions still evolving -- far from reaching a steady state -- by the end of the simulation time, providing further evidence that the critical state of the NWN occurs during the period that $x$ is actively changing. This result confirms how the dynamical process and activity of memristive junctions are important ingredients to maintain the power law characteristics in the frequency-time domain.

\begin{figure}[H]
    \centering
    \begin{subfigure}[b]{\linewidth}
        \centering
        \includegraphics[width=0.6\linewidth]{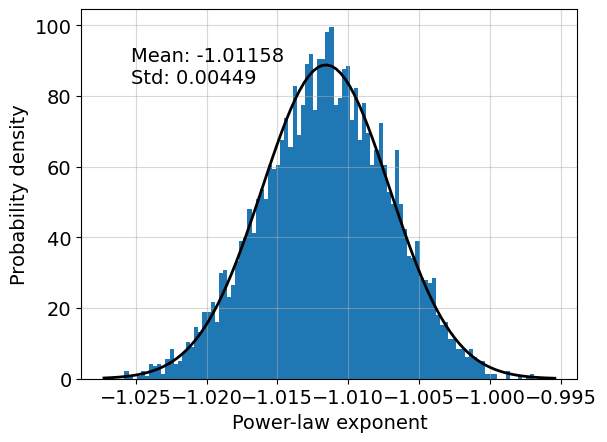}
        \caption{Ensemble using the Decay HP model. A bin size of 2.890 $\times 10^{-4}$ is used.}
        \label{fig:power-law-histogram}
    \end{subfigure}
    \vskip\baselineskip
    \begin{subfigure}[b]{\linewidth}
        \centering
        \includegraphics[width=0.6\linewidth]{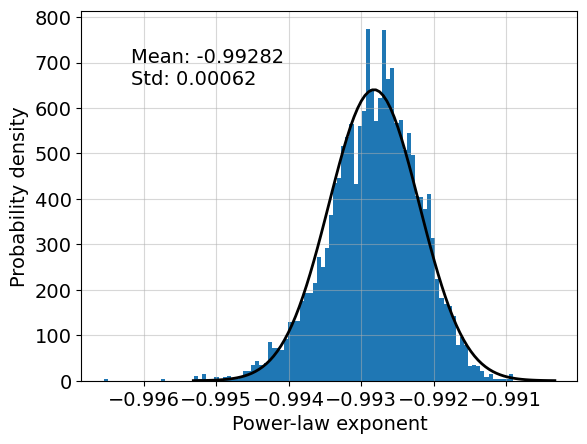}
        \caption{Ensemble using the SLT HP model. A bin size of 5.640 $\times 10^{-5}$ is used.}
        \label{fig:power-law-histogram-chen}
    \end{subfigure}
    \caption{Probability density histograms of power law exponents for an ensemble of 5000 simulated NWNs following the JDA representation. Each NWN sample in an ensemble was regenerated by randomly shuffling the nanowires spatially, but keeping the original wire density constant. A Gaussian function (solid black line) is overlaid onto each histogram using the mean and standard deviation taken from the exponent data. These values are shown in their respective panels.}
\end{figure}

To compare the occurrence of the power law trend across many randomly generated NWNs, an ensemble of 5000 NWNs following the JDA representation was created to perform the same simulations with both the Decay HP and SLT HP models. The fitted exponents for each of the NWNs were collected and graphed as an occurrence histogram distribution for the Decay HP model in \autoref{fig:power-law-histogram} and for the SLT HP model in \autoref{fig:power-law-histogram-chen}. On top of the distributions, we overlaid a Gaussian distribution using the sample mean and standard deviation taken from the respective distribution of power law exponents. These results reveal how much the $\beta$ exponent fluctuates with respect to the spectral analysis shown in \autoref{fig:I-vs-t} and \autoref{fig:I-vs-t-chen}. For the Decay HP model, the exponent $\beta$ extracted from \autoref{fig:I-vs-t} was $-\SI{1.0077\pm0.0008}{}$, which falls within one standard deviation of the mean for this simulated ensemble of NWNs, making the prior simulation a fairly typical output. Similarly, the exponent for the SLT HP model extracted from \autoref{fig:I-vs-t-chen} was $-\SI{0.993\pm0.001}{}$, which also places that simulation within one standard deviation of the mean of the SLT HP model power law distribution obtained with the ensemble. It should be noted that the Gaussian distributions plotted are not a fit of the data but rather a reference given the same mean and standard deviation taken from the data. The Decay HP model $\beta$ values appear to closely follow the reference Gaussian distribution, with the SLT HP model $\beta$ values similarly distributed with a slight left-skew deviation. Similar power-law behaviour was found in the current responses of NWNs at DC stimulation in our previous work in O'Callagan \textit{et al.} \cite{ocallaghan2018collective}, where we found $\beta\approx -1$, supported by experimental evidence. This suggests that the NWNs studied theoretically and experimentally in \cite{ocallaghan2018collective} follow closer to the dynamics and parameter phase space of the Decay HP model than the SLT HP model.

It is also important to compare our predictions for $\beta$ with earlier power spectral density (PSD) and $1/f$-power law dynamics analysis conducted in \cite{avizienis2012,sillin2013theoretical,dunham2021} and a more recent one in \cite{diaz-alvarez2019}. Overall, their predictions, which inspired our current studies, point to $|\beta|>1$, ranging from $|\beta|\sim 1.3$ up to $|\beta|\sim 1.7$. A first explanation for the discrepancy is that $|\beta|$ is likely sample-dependent; indeed, real-world NWNs can differ in numerous chemical and physical aspects, spanning from their connectivity and network morphology, as highlighted in \cite{dunham2021}, to distinct material/chemical compositions and electrochemical properties, and inherent stochasticity \cite{milano2025} in memristive phenomenon. Yet, the work of Diaz-Alvarez et al. \cite{diaz-alvarez2019} discusses an important feature that significantly impacts the value of $|\beta|$ in their measurements; they state ``when the network is in a dynamical regime that is stable against fluctuations, $\beta$ values are closer to 1, whereas they approach 2 when the network succumbs to instability''. Indeed, we did not introduce any source of instability in our DC-probed current simulations, such as junctions or nanowires suddenly breaking down to inflict an abrupt fluctuation in the equivalent conductance of the network. The current increases monotonically up to (near) saturation, and its Fourier transform clearly reveals $1/f$-power law dynamics with $|\beta|\approx 1$ for both modelling schemes, the Decay HP and the SLT HP, in agreement with Diaz-Alvarez et al. \cite{diaz-alvarez2019} interpretation.

\begin{figure}[H]
    \centering
    \begin{subfigure}[b]{0.475\linewidth}
        \centering
        \includegraphics[width=\textwidth]{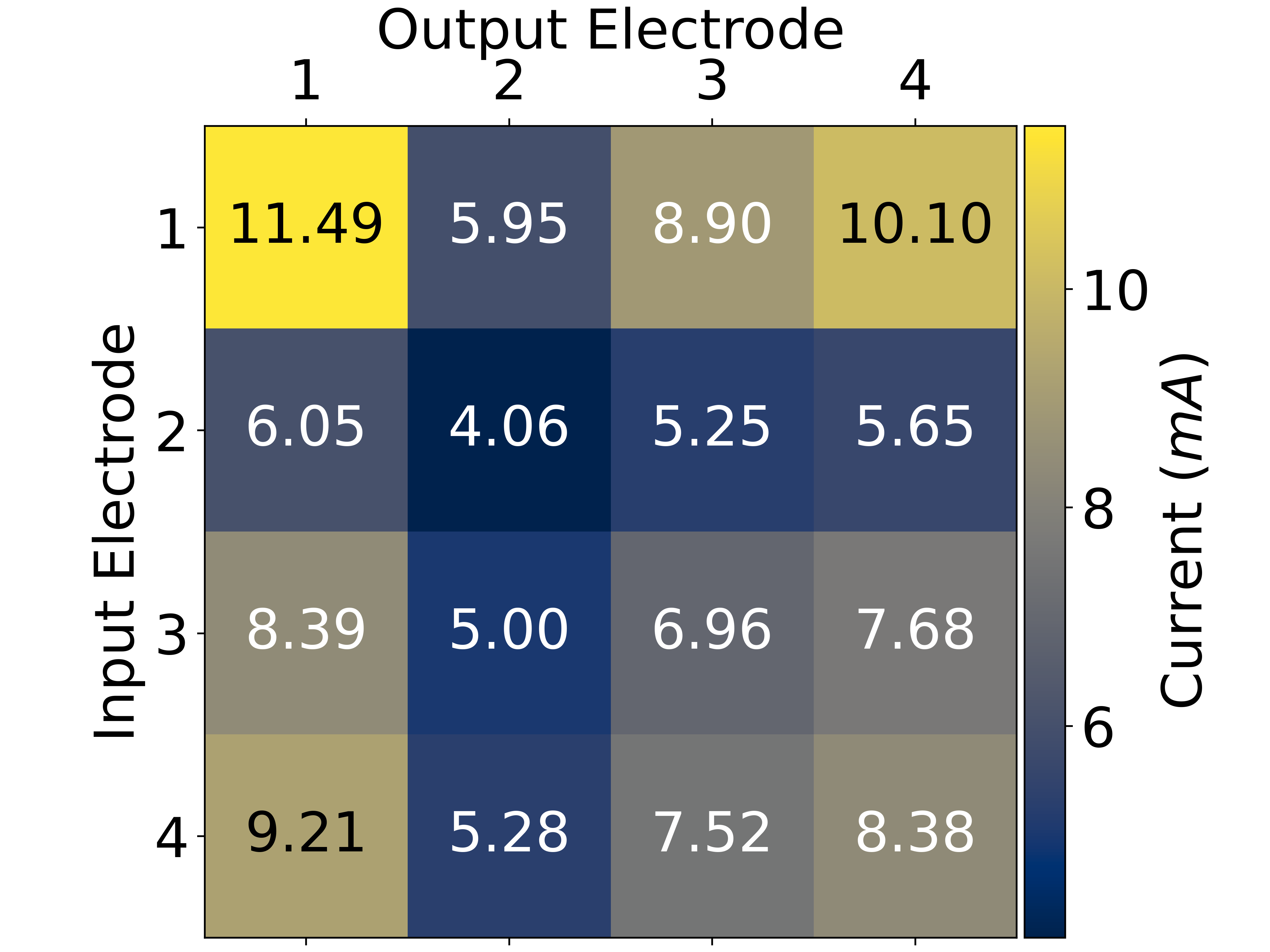}
        \caption{$\tau = \SI{1}{s}$ and a training time of \SI{10}{s}.}   
        \label{fig:multiple-decay100-max1000}
    \end{subfigure}
    \hfill
    \begin{subfigure}[b]{0.475\linewidth}  
        \centering 
        \includegraphics[width=\textwidth]{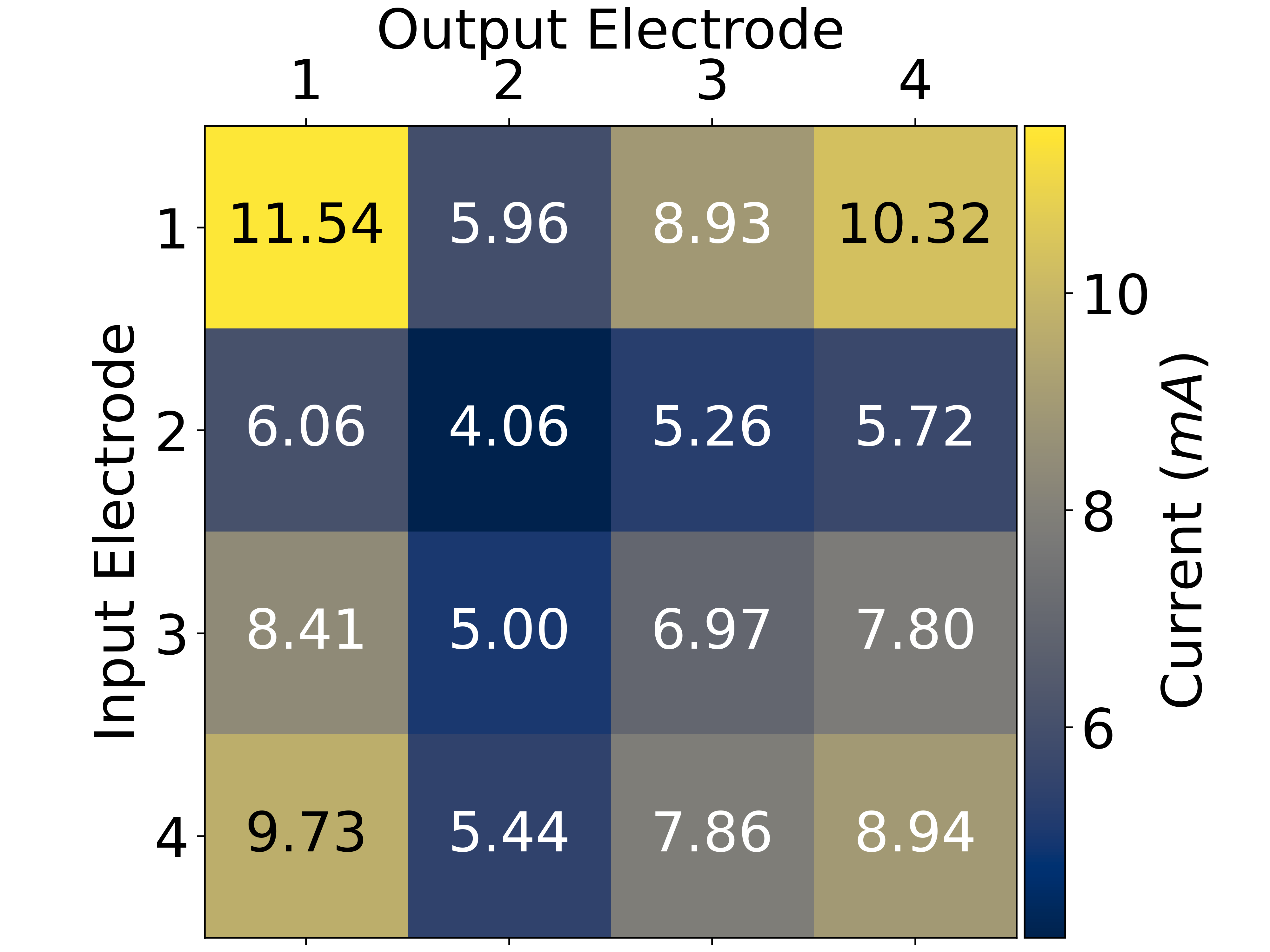}
        \caption{$\tau = \SI{1}{s}$ and a training time of \SI{50}{s}.}    
        \label{fig:multiple-decay100-max5000}
    \end{subfigure}
    \vskip\baselineskip
    \begin{subfigure}[b]{0.475\linewidth}   
        \centering 
        \includegraphics[width=\textwidth]{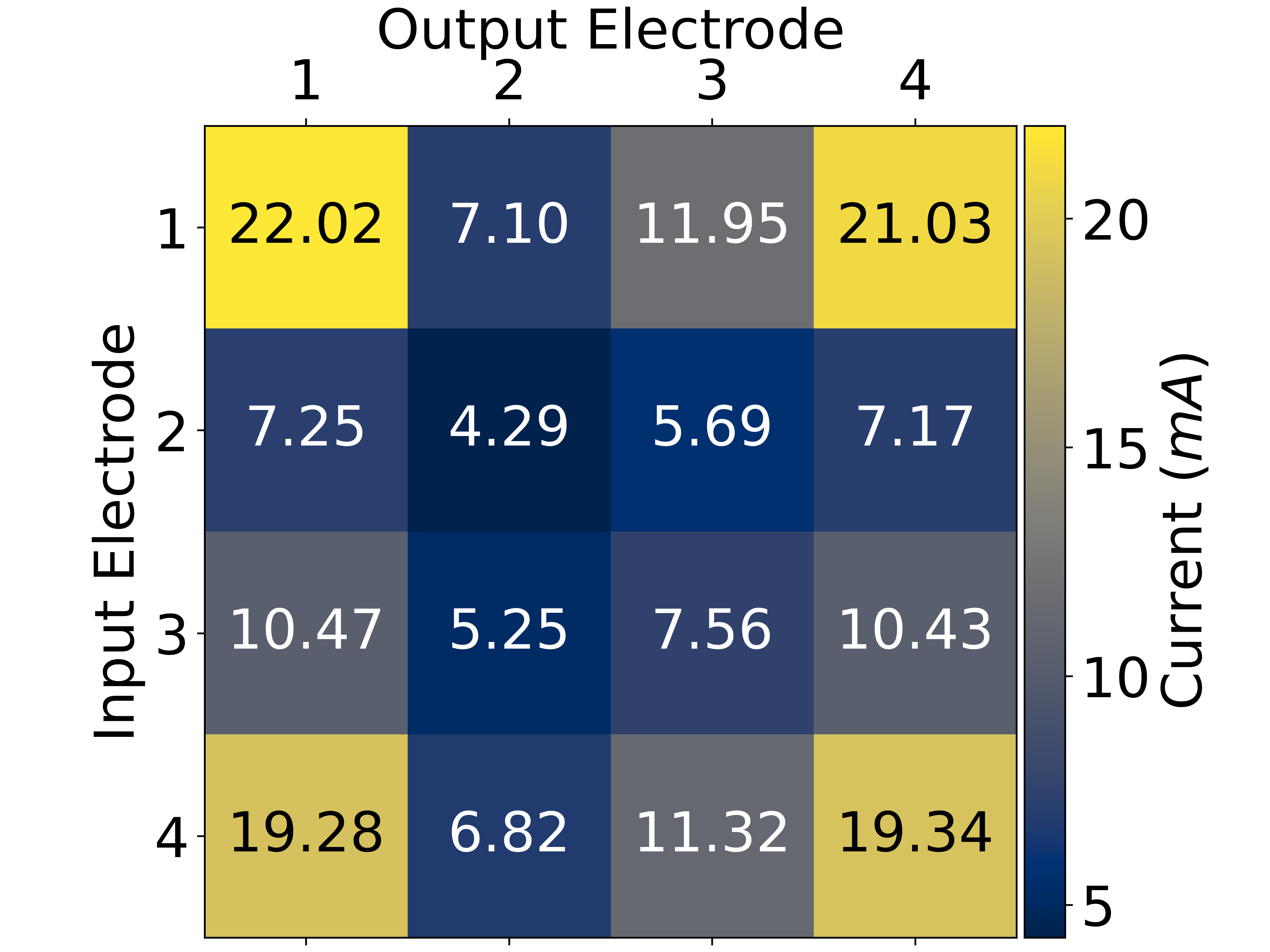}
        \caption{$\tau = \SI{2}{s}$ and a training time of \SI{10}{s}.}   
        \label{fig:multiple-decay200-max1000}
    \end{subfigure}
    \hfill
    \begin{subfigure}[b]{0.475\linewidth}   
        \centering 
        \includegraphics[width=\textwidth]{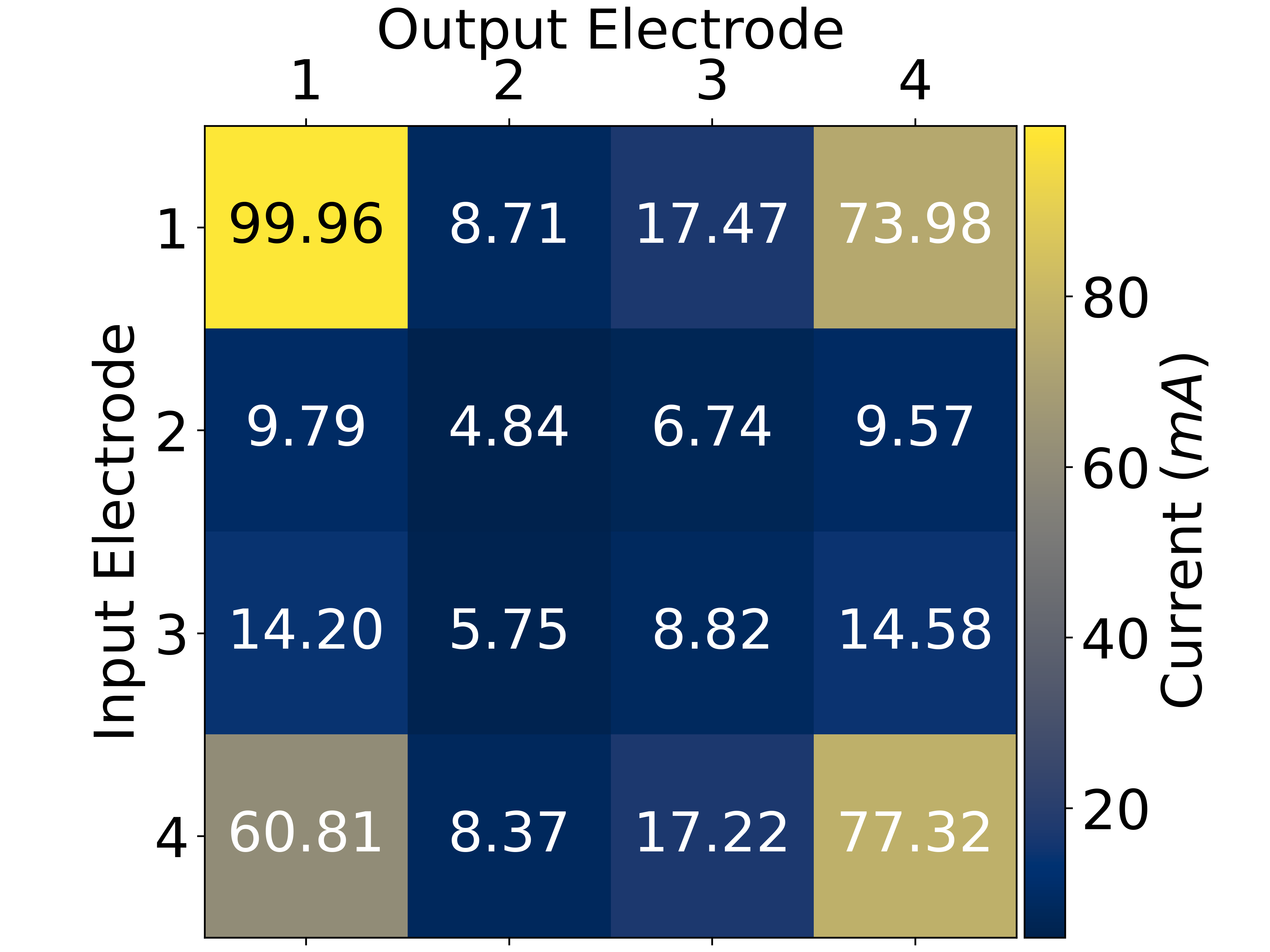}
        \caption{$\tau = \SI{2}{s}$ and a training time of \SI{50}{s}.}
        \label{fig:multiple-decay200-max5000}
    \end{subfigure}
    \caption{Confusion matrix of current flow readout taken post-training from each pair of input (IE) and output (OE) electrodes in a NWN system placed on a multi-electrode configuration and modelled by the Decay HP dynamical equations. The NWN has four electrodes on the left and four electrodes on its right-hand side. The readout current was taken after a training time consisting of applying a DC \SI{20}{V} stimulus on a pre-selected configuration of electrodes; in this case, IEs 1 and 4 are sources, whereas OEs 1 and 4 are grounded. All other unused electrodes are left floating during training. The decay rate ($\tau$) and training times are detailed below each panel.}
    \label{fig:multiple-electrodes}
\end{figure}

Following the investigation of the NWN dynamics subjected to a two-terminal setup, we move forward in enabling \texttt{MemNNetSim} to stimulate the network electrically via a multi-electrode architecture as proposed in a number of experimental realizations, for instance, in \cite{diaz-alvarez2020associative,milano2020}.
Here, we keep the same NWN characteristics as in our previous results, but set in the simulation four equally spaced electrodes on the left and right sides of the network. The left electrodes are classified as the input electrodes (IE) and the right electrodes as the output electrodes (OE). We start by simulating and quantifying current flow through this multi-terminal circuit configuration using the Decay HP model. In this scheme, we first ``train'' the NWN by applying a DC voltage onto pre-selected electrode pairs long enough to switch ON most of its junctions. After this training phase, a readout voltage difference is applied across each (IE,OE) pair, and its readout current can be organized in a 4$\times$4 confusion matrix representation. \autoref{fig:multiple-electrodes} illustrates current flow readout through the NWN when applying a voltage difference across each (IE,OE) pair, probed individually after training. The training was done by sourcing \SI{20}{V} (DC) onto IEs 1 and 4, and grounding OEs 1 and 4 for a sufficiently long time to saturate most of the NWN junctions. All other unused electrodes are left floating during training. With a decay rate of $\tau = \SI{1}{s}$ and a training time of \SI{10}{s}, we can see that the current flow from IE 1 is the highest between OEs 1 and 4; however, the current flow from IE 1 to OEs 2 and 3 is still substantial, despite those pathway domains not being explicitly trained. The same can be seen for IE 4. Even after a longer training time of $\SI{50}{s}$, this result is still observed, indicating the NWN cannot effectively be trained any further, and its structural pathways are not significantly contrasting. If all current values in the matrix were normalized, and a threshold of 0.5 is set to convert them into bits, this system could be seen as a bit pattern storage unit, which can store electrical information and can memorize ``by association'' if more than one pair of electrodes is used during training, as done in \cite{diaz-alvarez2020associative}. Nonetheless, the NWN studied in \autoref{fig:multiple-decay100-max1000} and \autoref{fig:multiple-decay100-max5000} did not perform such memorization task very well since the current pattern stored across the matrix is not well distinguished. In other words, since the training was done using IEs 1 and 4 and OEs 1 and 4, the expectation was to observe distinguishing current values at the corners of the confusion matrix, revealing that pathway domains directly connecting (IE=1,OE=1) and (IE=4,OE=4) would be significantly evolved, as well as cross pathways bridging (IE=1,OE=4) and vice versa. That is not the case in this example, therefore, we will try to improve this by increasing the decay rate of the NWN next. Yet, this result corroborates the phenomenon of heterosynaptic plasticity in NWNs as evidenced by \cite{milano2020}, in which the strengthening of non-directly-stimulated pathway domains is observed due to the recurrent complex connectivity, and self-organizing behaviour these systems exhibit.

A possible route to improve the imprint of differentiable information onto the network may be that a value of $\tau = \SI{1}{s}$ does not allow for sufficient junctions to be strengthened, resulting in pathways not being reinforced enough. On the bottom panels of \autoref{fig:multiple-electrodes}, we show the confusion matrices for the same NWN system, but with $\tau = \SI{2}{s}$ to observe the effects of an increased decay rate. Immediately, one can see on panel (c) that the current flow from IE 1 to OEs 1 and 4 is now much more contrasting in comparison to OEs 2 and 3, with similar results for IE 4. Allowing for further training emphasizes this improvement, as seen in \autoref{fig:multiple-decay200-max5000}. While the current flow of the untrained pathways does increase slightly with these parameters, the trained pathways in which the voltage difference was applied are clearly more distinct. In this way, we see the effect a larger decay constant has, along with the necessary training time in strengthening spatial pathways within the NWN.

\begin{figure}[H]
    \centering
    \begin{subfigure}[b]{\linewidth}  
        \centering
        \includegraphics[width=0.8\linewidth]{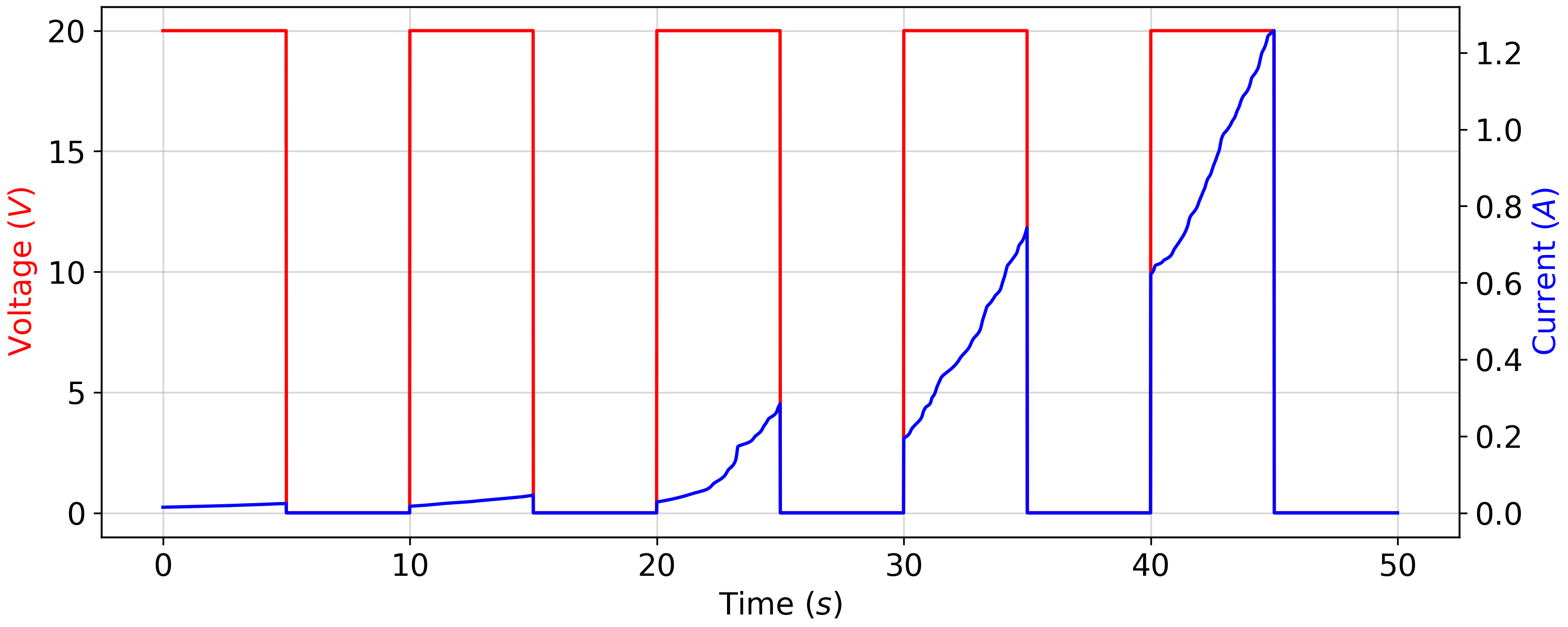}
        \caption{SLT HP model with parameters $\sigma=0.1$, $\theta=0.1$, and $a=0.8$, with initial conditions $x = 0.05$, $\tau=\SI{0.1}{s}$, $\varepsilon=0.1$.}
        \label{fig:temporal-memory-1}
    \end{subfigure}
    \begin{subfigure}[b]{\linewidth}
        \centering
        \includegraphics[width=0.8\linewidth]{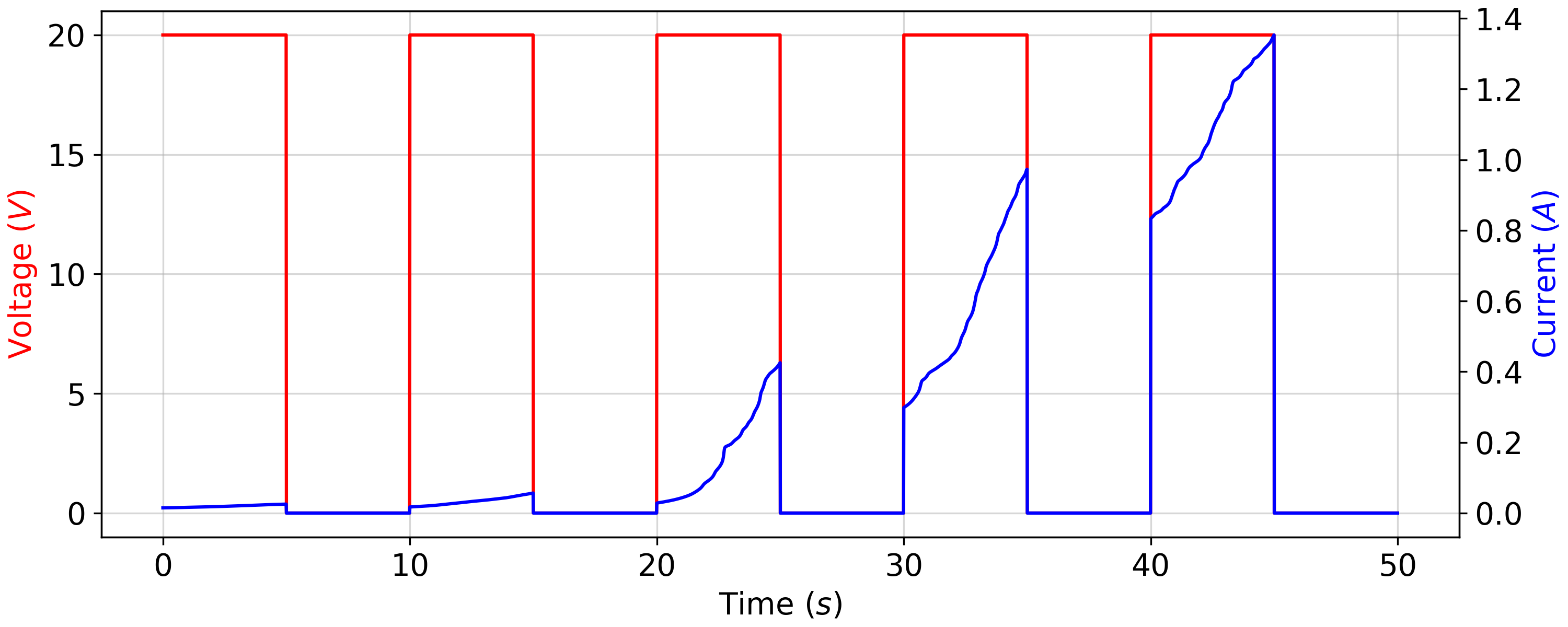}
        \caption{SLT HP model with parameters $\sigma=0.1$, $\theta=0.5$, and $a=0.8$, with initial conditions $x = 0.05$, $\tau=\SI{0.1}{s}$, $\varepsilon=0.1$.}
        \label{fig:temporal-memory-2}
    \end{subfigure}
    \begin{subfigure}[b]{\linewidth}  
        \centering
        \includegraphics[width=0.8\linewidth]{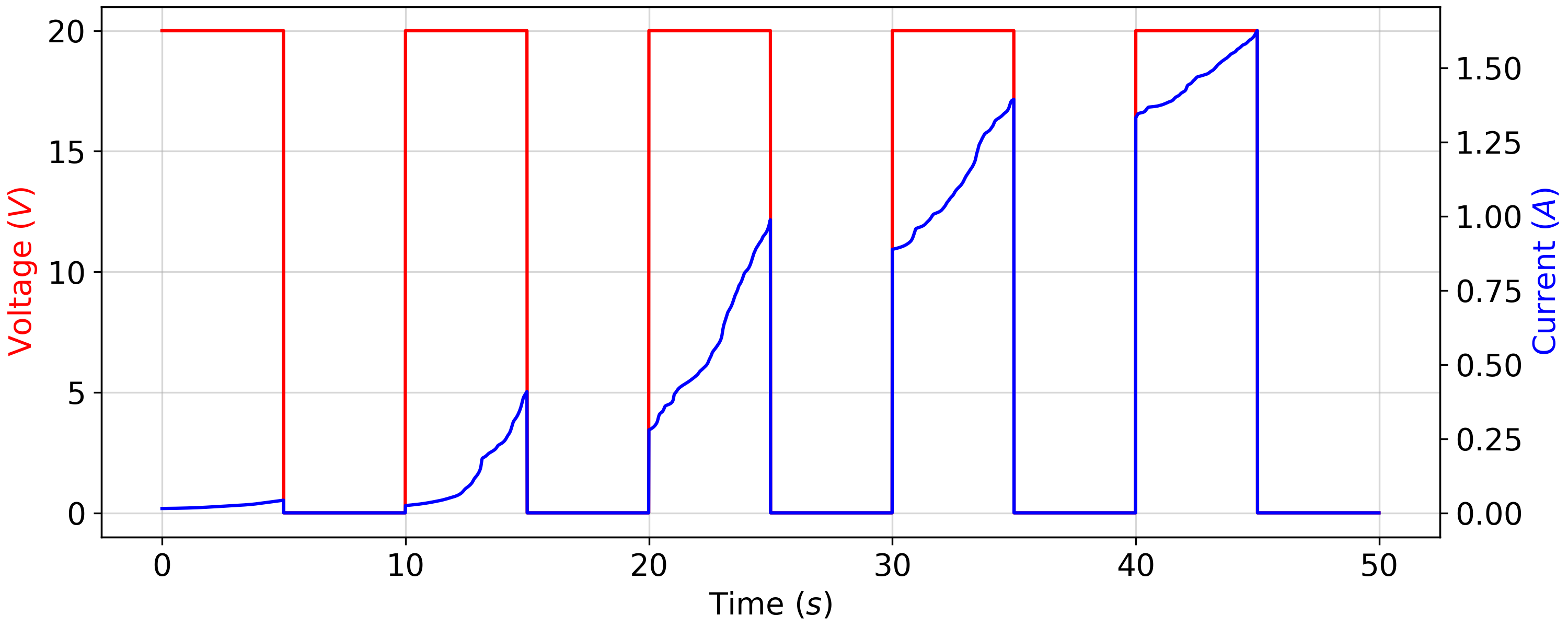}
        \caption{SLT HP model with parameters $\sigma=0.1$, $\theta=0.1$, and $a=0.8$, with initial conditions $x = 0.05$, $\tau=\SI{0.2}{s}$, $\varepsilon=0.1$.}
        \label{fig:temporal-memory-3}
    \end{subfigure}
    \caption{Voltage and current plots versus time for JDA NWNs using the SLT HP model with different model parameters and initial state variable conditions.}
    \label{fig:chen-model-profiles}
\end{figure}

Complementing the spatial analysis done previously, further simulations were performed using the SLT HP model to explore temporal dynamics in the form of current potentiation upon repeated voltage pulsing. This dynamics resembles paired-pulse facilitation (PPF) in synaptic plasticity. To do this, the NWN is subjected to voltage pulses of \SI{20}{V} for $\SI{50}{s}$ with interpulse intervals of $\SI{5}{s}$. \autoref{fig:chen-model-profiles} displays the applied voltage and output current flow across the NWN as a function of time for distinct parameterizations in the SLT HP model. In all cases, we observe the current increasing at each pulsation, intercalated by idleness phases in the intervals of zero voltage. It is worth mentioning that we do not observe a gradual relaxation of the current at zero bias because the models do not account for any capacitive feature in the junctions; they are purely resistive in nature and follow Ohm's law $V=RI$ in their response. Notably, it is interesting to infer the impact of the initial conditions for $\tau$ on the initial activation of the network. Little current activity is observed for the first two voltage pulses for the cases with $\tau(t=0)=\SI{0.1}{s}$, indicating the NWN at high resistance state. The current ramps up significantly after the third pulse, indicating the gradual formation of low resistance pathways within the NWN. Setting the junctions to initiate at a larger decay rate of $\tau(t=0)=\SI{0.2}{s}$ yields significant current activation earlier on (at the second voltage pulse), with the current ramping up at each subsequent pulse. In other words, the NWN was able to activate much quicker than in the case of $\tau(t=0)=\SI{0.1}{s}$, and, as a consequence, it was able to reach higher current values at the end of the voltage pulse sequence. $\theta$ is also a parameter that affects the evolution of $\tau$ in time, and it was also modified to study its influence (see equations \eqref{eq:chen-model}). By changing the parameter $\theta$ from $0.1$ to $0.5$, the effects of the decay rate changing can be seen in \autoref{fig:temporal-memory-2}. By increasing $\theta$, the decay rate variable is able to more effectively change with response to the input voltage, leading to more prominent current increases at the beginning of the pulse and higher current values after the five voltage pulses.

\begin{figure}[H]
    \centering
    \includegraphics[width=0.7\linewidth]{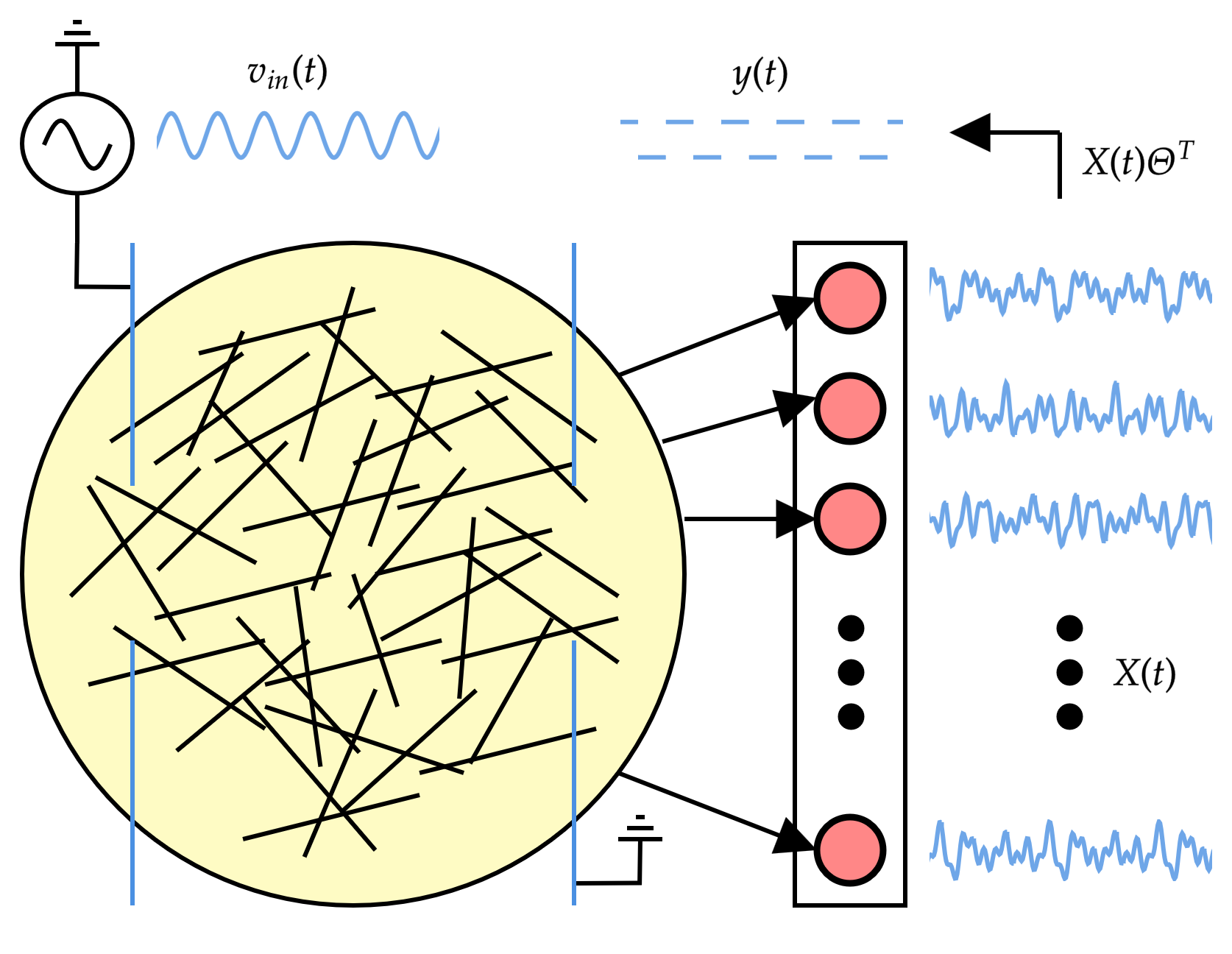}
    \caption{Schematic of a NWN used as a physical reservoir in nonlinear wave transformation. The NWN represented by the randomized black sticks is subjected to a sinusoidal input voltage, $v_{in}(t)$,  sourced at opposite-side electrodes (represented by blue vertical sticks). The reservoir output is passed to a readout layer, which has its weights ($\Theta$, represented by the black arrows) optimized following MSE minimization given a target time-series (periodic) pattern, $T(t)$. During training, the dynamical internal state of the reservoir, $X(t)$, evolves in time, and the readout weights are optimized, outputting the transformed signal $y(t)=X(t)\Theta^T$. Training phase is set to be completed when most of the NWN junctions have nearly saturated to the LRS with $y(t)\rightarrow T(t)$, and the readout weight optimization can be interrupted.}
    \label{fig:reservoir}
\end{figure}

Reservoir computing with NWNs has been popularized for its adaptable abilities to process information, classify data, and for predictive tasks \cite{fu2020reservoir,moon2019temporal,du2017reservoir}. Using the framework provided in \texttt{MemNNetSim} as a foundation, we are able to simulate memristive-based reservoir computing (RC) in random NWNs with more flexibility and customization settings. We focused on reservoir architectures that are inspired by the so-called {\it in-materio} RC paradigm as presented in \cite{milano2022v2,fang2023,kotooka2024}. In this framework, the NWN acts as a physical reservoir that projects low-dimensional input information into a high-dimensional space to which is then passed on and processed by a readout layer as depicted in the \autoref{fig:reservoir} schematics. This is an open-loop architecture on which the NWN conducts a nonlinear mapping of a time-dependent input signal that will be ``imprinted'' into its internal state, accessed by the readout layer. We will employ this scheme to the specific application of waveform transformation operation in which the RC device receives a relatively simple signal input and learns how to reproduce a more complicated signal pattern based on a target function. The error between the learned output and the target is monitored at the readout layer, in which its output weights are optimized during the training phase to minimize a mean-square error loss function. It is important to note that in this particular RC architecture, the reservoir has dynamic weights since the memristive junctions do evolve with time. As discussed in \cite{tanaka2019recent,fang2023}, the essential dynamical properties that impact RC performance are (i) short-term memory, (ii) nonlinearity, (iii) separation, (iv) approximation, and (v) high-dimensionality. With a sufficiently high nanowire density or network size, the high dimensionality requirement can be satisfied; however, the other properties will depend on the specific memristive model chosen. Here, we will analyze the effects and RC performance for nonlinear wave transformation NWN-based devices, specifically using the HP and the Decay HP model, so we can directly assess the existence (or not) of short-term memory property in the interwire junction model. In the following, we will explain in more detail the mathematical framework used to implement such physical RC scheme that we adapted from \cite{fu2020reservoir}.

To start, a JDA NWN was generated with four electrodes (two on its left and two on its right-hand side) and selected one left electrode as the source and another as the drain (right side), with the other electrodes left floating. The voltage of all $N$ nanowires was used to represent the reservoir state. The NWN was then trained for a period of time by stressing it with a sinusoidal voltage input. During this time, the reservoir state was denoted as and recorded in the variable set
\begin{align}
    X(t) = \begin{bmatrix}
        v_1(t), v_2(t), v_3(t), \cdots, v_N(t)
    \end{bmatrix}
\end{align}
where $t$ is the time, and $v(t)$ is the voltage for each particular nanowire seen as an isopotential nodal entity within JDA picture, and that can be recorded for a total of $M$ time steps. The reservoir state was then used to train the weights of an output readout layer through linear regression to produce a desired target signal $T(t)$. To do this, the mean-squared error (MSE) was used as the loss function,

\begin{align}
    J(\Theta) = \frac{1}{M} \sum_{m=1}^M (T(t_m) - y(t_m))^2
\end{align}
with $y(t) = X(t)\Theta^T$ as the output signal and the readout layer weights as
\begin{align}
    \Theta = \begin{bmatrix}
        \Theta_1, \Theta_2, \Theta_3, \cdots, \Theta_N \, 
    \end{bmatrix}
\end{align}
with $\Theta^T$ being its transpose. To minimize the MSE, the Broyden-Fletcher-Goldfarb-Shanno (BFGS) quasi-Newton algorithm implemented in \texttt{SciPy} \cite{pauli2020scipy} was used with the MSE gradient given by
\begin{align}
    \frac{\partial J(\Theta)}{\partial \Theta_n} = -\frac{2}{M} \sum_{m=1}^M (T(t_m) - y(t_m)) v_n(t_m) \, .
\end{align}
\noindent with $n=1,\dots,N$. For this work, a sinusoidal input voltage was selected with a DC offset of the form
\begin{align}
    v_\text{in}(t) = \frac{V_0}{2}\brac{\sin(2\pi ft) + 1}
    \label{eq:input_voltage}
\end{align}
where $V_0 = \SI{20}{V}$ and $f = \SI{5}{Hz}$ along with a square wave target signal
\begin{align}
    T(t) = \frac{V_0}{2}\brac{\sgn(\sin(2\pi ft)) + 1}
\end{align}
Finally, Joglekar's window function \cite{joglekar2009} with $p = 3$ was chosen for the memristive state equations simulations. Once more, \autoref{fig:reservoir} illustrates the general schematic for a NWN as a physical reservoir for the example of the nonlinear wave transformation. Since computationally all the voltages of the nanowires in the NWN are known, the readout layer is assumed to connect to every nanowire for use in its weight optimization. In practice, this may not be feasible; however, multiple gate probes, for instance, incorporated at the bottom of the reservoir area, could act as proxies for groups of nanowires to then be connected to a readout layer instead. In this way, the desired readout layer training could still be achieved even though some signal degrees of freedom will be inherently lost since the macro-gates will essentially ``bin'' the output signal from nanowire clusters to the readout layer.

\begin{figure}[H]
    \centering
    \includegraphics[width=0.8\linewidth]{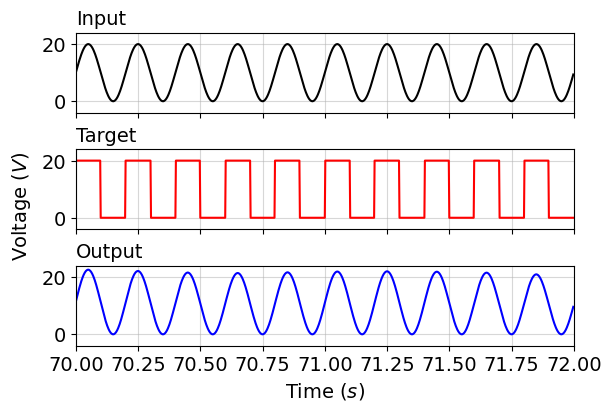}
    \caption{Input, target, and output waveform in the nonlinear wave transformation process using a NWN as a physical reservoir with each nanowire junction ruled by the HP Model, during the training period of the readout layer. The RNMSE for this duration was 0.3275.}
    \label{fig:wave-transformation-hp}
\end{figure}

\begin{figure}[H]
    \centering
    \includegraphics[width=0.8\linewidth]{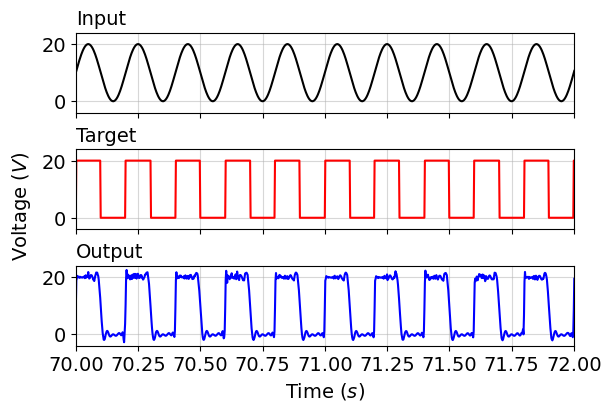}
    \caption{Input, target, and output waveform in the nonlinear wave transformation process using a NWN as a physical reservoir with each nanowire junction ruled by the Decay HP Model, during the training period of the readout layer. The RNMSE for this duration was 0.1410.}
    \label{fig:wave-transformation}
\end{figure}

\begin{figure}[H]
    \centering
    \begin{subfigure}[b]{\linewidth}
        \centering
        \includegraphics[width=0.8\textwidth]{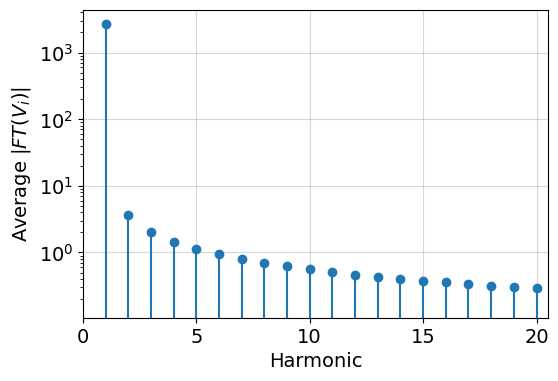}
        \caption{HP Model in the memristive junctions.}
        \label{fig:harmonics-HP}
    \end{subfigure}
    \vskip\baselineskip
    \begin{subfigure}[b]{\linewidth}
        \centering
        \includegraphics[width=0.8\linewidth]{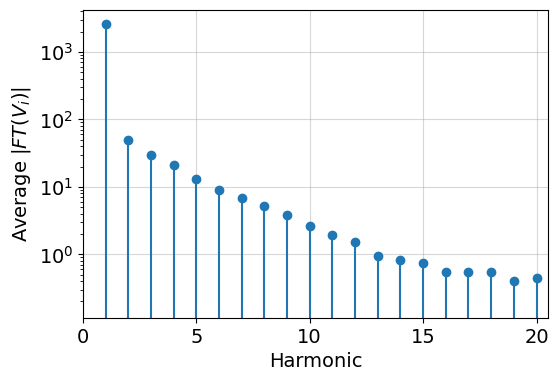}
        \caption{Decay HP Model in the memristive junctions.}
        \label{fig:harmonics}
    \end{subfigure}
    \caption{Average harmonic strength of Fourier transformed voltage traces of a NWN reservoir trained from 70 s to 72 s using the HP model (top panel) and Decay HP model (bottom panel) in the memristive junctions.}
\end{figure}

Our first simulation considers the RC scheme as in \autoref{fig:reservoir} and the interwire junctions ruled by the ionic linear drift HP model. The initial conditions for all junctions are set at $x(t=0) = 0.05$ and the NWN was driven by the input signal in \autoref{eq:input_voltage} for \SI{100}{s}. The readout layer was trained during the interval of \SI{70}{s} to \SI{72}{s}. \autoref{fig:wave-transformation-hp} shows the sinusoidal input signal alongside the square wave target signal and the resultant output signal after application of the trained readout layer. Despite reservoir computing requirements of high dimensionality and nonlinearity being satisfied, it is visually apparent that the output signal does not match the target signal and was unsuccessful in its task to transform the sine wave into a square wave. This confirms that the absence of short-term memory properties in the HP model does not allow for a successful RC wave transformation. The system did not really learn anything; it just kept driving/copying the input signal. Quantitatively, the success of the signal production can be measured in a given region of time with the root-normalized mean-squared error (RNMSE)
\begin{align}
    \text{RNMSE} = \sqrt{\frac{\sum_{m=1}^M (T(t_m) - y(t_m))^2}{\sum_{m=1}^M (T(t_m))^2}}
\end{align}
with sufficiently low RNMSE indicating better wave transformation performance. For the HP Model, a RNMSE of 0.3275 was found, and we will verify if this value can be reduced when the memristive junction model is able to account for short-term memory. As a result, next we will adopt the Decay HP model and test the NWN ability to transform a sine wave into a square wave. The NWN parameters were kept the same as in previous simulations, with the addition of the decay rate set at $\tau = \SI{1}{s}$. The result is seen in \autoref{fig:wave-transformation}, during the training period of the readout layer. Despite some noise and characteristic Gibbs phenomenon features, the NWN successfully transformed the sinusoidal input into a square wave output, resembling the target function. The computed RNMSE in this case is 0.1410, revealing in quantitative terms that this transformation is more successful than the one attempted by the NWN under the HP model. However, why exactly is the NWN under the HP model incapable of nonlinearly transforming the sinusoidal input into the square wave time series? To understand that, we took the Fourier transform of all nanowires' time-dependent voltage outputs, and conducted the pointwise mean of all nodal power spectra to condense into a single (averaged) power spectrum plot. Further, integer multiples of the input signal frequency can be selected to represent the NWN harmonics. \autoref{fig:harmonics-HP} provides reasoning for the HP model's inability to perform the wave transformation task. The strength of the higher harmonics is around three magnitudes weaker than the fundamental frequency and cannot effectively contribute to the target wave construction. On the other hand, the Decay HP model harmonics in \autoref{fig:harmonics} have higher harmonics which are much stronger in comparison to the fundamental frequency. This means that junctions under the Decay HP model are able to ``break down'' the input signal into a more versatile harmonic profile that can reconstruct the square wave pattern. The short-term memory property allows the network to adapt and learn the new pattern by ``forgetting'' redundancies or unnecessary information from the input signal. This phenomenon is similar to the high-harmonic generation (HHG) capabilities observed experimentally in Ag$_2$Se NWNs by Kotooka et al. \cite{kotooka2024} and other nanoscale network systems such as Ag-Ag$_2$S nanoparticles studied in \cite{hadiyawarman2021,srikimkaew2024}. Waveform classification tasks in memristive RC NWNs were also successfully demonstrated experimentally by Weng et al. \cite{weng2024} for distinct waveform patterns, including temporal patterns with combined square-triangular pulses, in which the authors also argue the ``learning-forgetting-relearning'' process that the network goes through to perform the transformation. In this way, we see evidence that supports the short-term -- fading memory -- requirement to use NWNs as a reservoir.

\begin{figure}[H]
    \centering
    \includegraphics[width=0.8\linewidth]{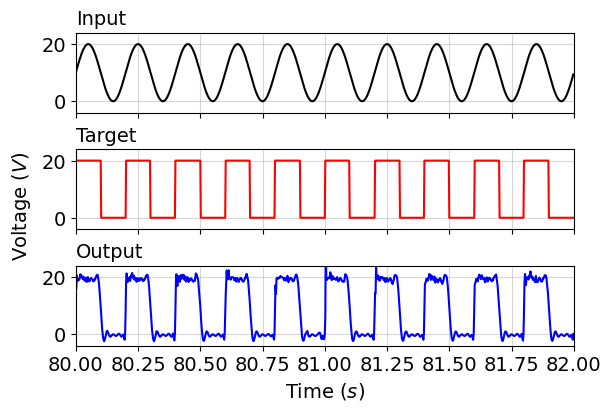}
    \caption{Input, target, and output waveform in the nonlinear wave transformation process using a NWN as a physical reservoir with each nanowire junction ruled by the Decay HP Model, after the training period of the readout layer. The RNMSE for this duration was 0.1557.}
    \label{fig:wave-transformation-testing}
\end{figure}

An important investigation that needs to be carried out is to assess how robust the training of the readout layer weights is and if the NWN RC device can continue reproducing the target waveform pattern for long periods of time, even if the readout layer weights have been fixed at an optimum state. \autoref{fig:wave-transformation-testing} illustrates the same NWN as in \autoref{fig:wave-transformation} using the Decay HP model for the memristive junctions and fixing the readout layer weights at their optimum state after the training period from \SI{70}{s} to \SI{72}{s}. From the displayed time period from \SI{80}{s} to \SI{82}{s}, we can see that the input signal is nonlinearly transformed to the intended square waveform. The RNMSE for this period was computed to be slightly enhanced to 0.1557; yet, the NWN RC device successfully performed the transformation.

\begin{figure}[H]
    \centering
    \begin{subfigure}[b]{\linewidth}
        \centering
        \includegraphics[width=0.8\linewidth]{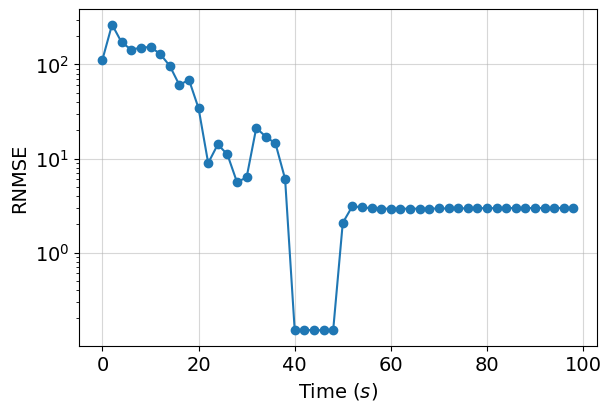}
        \caption{Readout layer training between 40 s and 50 s.}
        \label{fig:rnmse-vs-time-early}
    \end{subfigure}
    \vskip\baselineskip
    \begin{subfigure}[b]{\linewidth}
        \centering
        \includegraphics[width=0.8\linewidth]{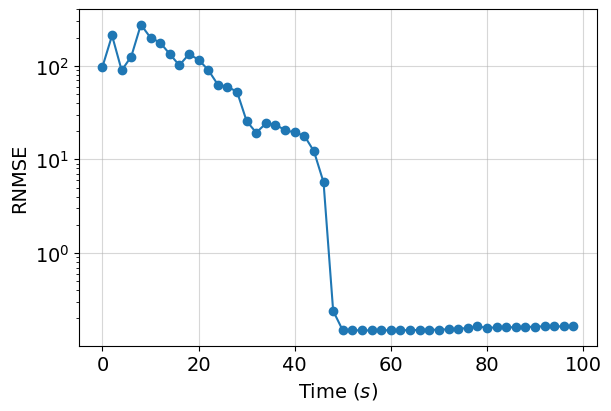}
        \caption{Readout layer training between 50 s and 60 s.}
        \label{fig:rnmse-vs-time}
    \end{subfigure}
    \caption{RNMSE as a function of time for a JDA NWN using the Decay HP model with $\tau = 1$ s. RNMSE values are calculated in increments of 2 s.}
    \label{fig:rnmse-vs-time-compare}
\end{figure}

There is a notable difference in nonlinear wave transformation performance depending on when (and for how long) the readout layer is trained. Using the RNMSE calculated in time increments of \SI{2}{s}, the RNMSE can be plotted over time to gauge the transformation performance. Consider a training period from \SI{40}{s} to \SI{50}{s}; \autoref{fig:rnmse-vs-time-early} illustrates the RNMSE time evolution, in which we can observe its gradual reduction over time, achieving its lowest values during the training period. However, if the training period is somehow interrupted sufficiently early, the RNMSE value jumps back up, indicating the NWN junctions had not yet been settled into a steady state. This would also indicate an inferior waveform transformation. By allowing the junctions to evolve further and training the readout layer from \SI{50}{s} to \SI{60}{s} as seen instead in \autoref{fig:rnmse-vs-time}, the RNMSE after the readout layer training remains much closer to the lower baseline value set during this training, and it remains there. As a result, this evidences the NWN RC device acted a robust waveform transformation that will last for longer periods of time.

\begin{figure}[H]
    \centering
    \begin{subfigure}[b]{\linewidth}  
        \centering
        \includegraphics[width=0.8\linewidth]{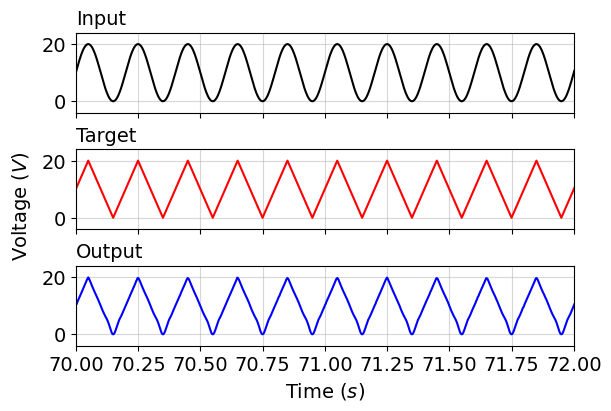}
        \label{fig:wave-transform-sine-triangle}
    \end{subfigure}
    \vskip\baselineskip
    \begin{subfigure}[b]{\linewidth}
        \centering
        \includegraphics[width=0.8\linewidth]{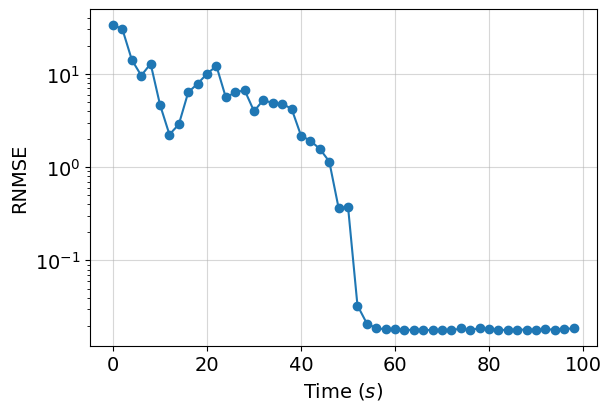}
        \label{fig:rnmse-sine-triangle}
    \end{subfigure}
    \caption{(Top panel) Input, target, and output waveform in the nonlinear wave transformation process using a JDA NWN as a physical reservoir with each nanowire junction ruled by the Decay HP Model, during the training period of the readout layer. The decay rate used was $\tau = 1$ s. The RNMSE for this duration was 0.0179. (Bottom panel) RNMSE as a function of time during the whole NWN evolution. RNMSE values are calculated in increments of 2 s. The achieved output pattern on the left panel was taken from the time interval between 70 s and 72 s.}
    \label{fig:sine-triangle}
\end{figure}

\begin{figure}[H]
    \centering
    \includegraphics[width=0.8\linewidth]{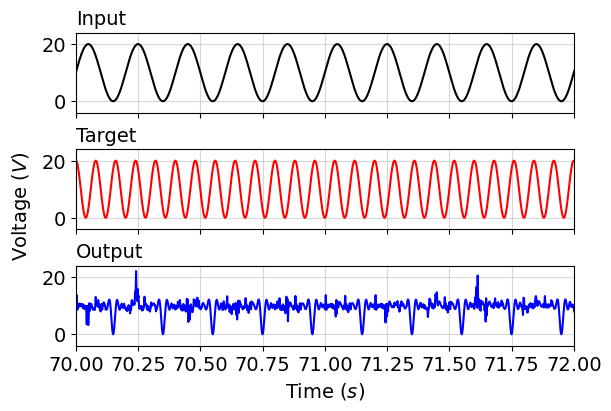}
    \caption{Input, target, and output waveform in the nonlinear wave transformation process using a NWN as a physical reservoir with each nanowire junction ruled by the Decay HP Model, during the training period of the readout layer. The target waveform is a frequency 2.5 times the input frequency. The RNMSE for this duration was 0.6012.}
    \label{fig:wave-transformation-non-int}
\end{figure}

To probe another wave-transformation task on the NWN RC under the Decay HP dynamics, we set a different target pattern, the triangular wave. The same sinusoidal wave as inputted and the NWN RC task is to convert it to a triangular waveform. This transformation is depicted in \autoref{fig:sine-triangle} as well as the evolution of the RNMSE during training. One can immediately notice that the reservoir executed this transformation with less noise and more seamlessly than the square wave. In this example, the target outlines the input function, and the transformation mostly involves mitigating the smoothness of the sine wave by working out the sharp triangular structures. Even the RNMSE already starts considerably lower than in the square wave training (one order of magnitude lower), evolving as well to even lower values (below $10^{-1}$) at similar threshold times (around 50 s).

One last interesting example we wish to showcase is a situation in which the NWN RC cannot reproduce a target, and we can explain this phenomenon through the harmonic spectral analysis conducted earlier. We passed to the network the input sine wave as in \eqref{eq:input_voltage}, and set as the target an inharmonic wave formed by 
\begin{align}
    T(t) = \frac{V_0}{2}(\sin(2\pi ft + \phi) + 1)
    \label{eq:inharmonic}
\end{align}
where $V_0 = \SI{20}{V}$, $f = \SI{12.5}{Hz}$, and $\phi = \pi/2$. The resulting target waveform is a frequency 2.5 times the input frequency. This waveform transformation attempt is shown in \autoref{fig:wave-transformation-non-int}, and one can see that the output does not reproduce the inharmonic oscillation. The output is a noisy baseline with periodic zero-voltage depths that synchronize with the valleys of the input sine wave. The failure is related to the fact that the NWN reservoir breaks down the input signal into harmonic modes and does not have any inharmonic modes at its disposal to recreate the target. Not shown here is the RNMSE evolution for this training, and one can expect large error values due to the clear discrepancy between the output and the target. The output indicates the NWN reached a steady state dynamics, supported by the background voltage fluctuations, intercalated by reset points that track the input frequency and could be used as switching events or time reference circuitry.

    \section{Conclusions}
This work is a journey through \texttt{MemNNetSim}, our comprehensive computational toolkit to model the static or dynamical behaviour of complex random NWNs. Within the static regime, the network is viewed purely as a resistive system in which the interwire contact points (or junctions) and the wire segments are modelled as simple Ohmic resistors according to the complete graph nodal representation, MNR. For NWNs in which the junction resistances are typically much larger than the wire inner resistances, the JDA can be adopted, in which the only source of resistance in the network comes from the interwire connections. The package enables both graph representations when mapping the NWN as a complex circuit using Kirchhoff's laws and MNA algorithm.

The more interesting aspect of \texttt{MemNNetSim} is not simply the package itself, but its ability to simulate emergent physical phenomena in NWNs, prone to neuromorphic response, when the dynamical feature is enabled. Within this picture, \texttt{MemNNetSim} considers the interwire junctions as memristive units with time-dependent resistance state. This allows the investigation of intelligent behaviour in random NWNs, starting from their memorization capability manifested in voltage versus current hysteresis loops and current potentiation upon a history of voltage excitation. The memristive models implemented in the package so far are the HP \cite{strukov2008missing}, the Decay HP \cite{sillin2013theoretical}, and the SLT HP \cite{chen2014phenomenological}; they differ in complexity in terms of their characterization of ionic migration in the memristive junctions, with the simplest picture from the HP model considering linear ion drift, up to the Decay and SLT constructions that can include short- and long-term memory effects in the evolution of the internal state equations. In the future, we plan to include nonlinear models that deviate from the linear input-output memristive form of $V(t) = R(t)I(t)$ to account for nonlinear drifts such as in quantum tunnelling, trap-assisted switching, and Schottky barrier effects.

With the three memristive descriptions already available in the package, we were able to simulate remarkable brain-like activity in the NWNs, with supporting evidence found in the literature. In particular, we simulated the current evolution of NWNs exposed to long periods of DC stimulation and performed a spectral analysis that evidenced the characteristic frequency power law also found in real experimental NWN samples. We obtained critical exponents ruling the scaling law $f^{-\beta}$ close to 1, indicating dynamical stability in the NWNs, corroborated by the explicit monitoring of all junction state variables and visualized over time. The package also supports NWN circuit design with multi-electrode architecture, from which we were able to demonstrate spatial associative memory through the emulation of a bit-grid storage pattern computed through the adaptive and selective current flow in the network. We were able to pinpoint that the decay rate $\tau$ can significantly affect the contrasting patterns, with the NWN performing superiorly at sufficiently larger $\tau$ values. Finally, we implemented an open-loop memristive-based reservoir computing scheme adapted from \cite{fu2020reservoir}, in which we successfully simulated the NWN acting as a physical reservoir for waveform transformation. In this scheme, the NWN was capable of transforming a low-dimensional voltage signal, e.g., a simple sinusoidal wave, into a higher-dimensional (more complicated) pattern, in our case, a square or a triangular wave with sharp edges. This was achieved by letting the NWN adapt in time while training the weights of a readout layer, idealized to be connected to its voltage nodal points. The training follows a least-square minimization against a desired target function. We demonstrated cases in which the transformation was successful, but also cases in which it failed to verify the best ``ingredients'' for transformation. We found that once again, the decay rate $\tau$ is an important factor for the NWN to learn the new pattern, in agreement with literature pointing out the importance of short-term memory in reservoir computing. That is one of the reasons that the simple linear drift HP model was not able to ``copy'' the target signal, whereas the Decay HP achieved the task. Our simulations also show that the effective reproduction of the waveform is related to the harmonic spectral profile obtained from the NWN during training; the NWN with spectral profiles with stronger higher harmonics was capable of conducting the wave transformation, evidencing its HHG capabilities. Besides the decay rate, the time for activation of the NWN and initiation of the training are important factors in the reservoir computing wave-transform. For instance, in our results, letting the NWN evolve for 10 seconds longer yielded better transformations and the permanence of the loss function at minimum values. For the shorter activation time, the loss function did reach a minimum, but it kicked back into higher values once the training time ended, indicating that the NWN missed the target pattern. One last failed attempt at wave transformation was presented to cycle back to the spectral harmonic analysis and HHG. We input a harmonic sinusoidal wave and set as the target an inharmonic signal with a fractional frequency component. In this example, the NWN reservoir processes and breaks down the input signal into harmonics, which are integer multiples of the fundamental frequency. As a result, there are no inharmonic modes at its disposal to reconstruct the target, resulting in a failed wave transformation. This failed attempt demonstrates the importance of assessing the spectral profile of the NWN reservoir to evaluate its capacity to replicate the target.

In the future, we plan to extend memristive-based reservoir computing strategies for the NWN to replicate non-periodic signals, establish network theory metrics to quantify how the NWN connectivity affects the waveform learning, incorporate capacitive effects in the junctions, as well as stochastic effects to account for thermal variations. We will also attempt to design and probe other bio-inspired learning methods, such as STDP, and pattern classification through unsupervised learning. This is just the beginning for \texttt{MemNNetSim}, which will continue expanding its simulation framework dedicated to neuromorphic NWN systems, so we can inform potential brain-inspired architectures that can be tested in laboratory.
    
    \section*{Acknowledgements}
    C.G.R. acknowledges that this publication has emanated from research supported by UCalgary Research, UCalgary Faculty of Science, the Natural Sciences and Engineering Research Council of Canada (NSERC) (USRA, Discovery Grant, and Alliance International Catalyst programs), the Quantum City initiative, and the NSERC Alliance - Alberta Innovates Advance Program (Streams I and II). This publication has also emanated from research supported in part by a research grant from Science Foundation Ireland (SFI) under Grant No. SFI/12/RC2778\_P2. W.N. was funded by an NSERC Discovery Grant (RGPIN/04568-2020), a Canada Research Chair (CRC-2019-00416), and the Hotchkiss Brain Institute. We also acknowledge the Advanced Research Computing (ARC) at UCalgary, the specialized UCalgary Customer Technology Services team of T. MacRae and B. Michaels, the Digital Research Alliance of Canada (former Compute Canada), and CMC Microsystems for computational resources. We thank C. Soriano for the design and creation of the \texttt{MemNNetSim} logo and banner.
    
    \balance
    
    \bibliography{rsc}
    \bibliographystyle{rsc}

    \newpage
    
\begin{center}
\LARGE{Supplementary Information: Characterizing Memristive Nanowire Network Models via a Unified Computational Framework}
\end{center}

\

\begin{center}
\large{M. Kasdorf, et al.}
\end{center}

\section*{Static Transport Analysis}

Besides dynamical memristive description in NWNs, \texttt{MemNetSim} can also provide static transport assessment. This includes calculation of the equivalent sheet resistance of the network, voltage nodal information, and current distribution for fixed junction and inner wire resistances, i.e., without the memristive time evolution. This is set by disabling the dynamical simulation workflow in the package. Static transport analysis was extremely useful to benchmark the package by retrieving well-known dependencies in (Ohmic) resistive NWNs that confirmed the correct implementation of the Kirchhoff's circuit laws via MNA, graph representation of the network, and large-ensemble analysis at static regime. 

\begin{figure}[H]
    \centering
    \includegraphics[width=0.8\linewidth]{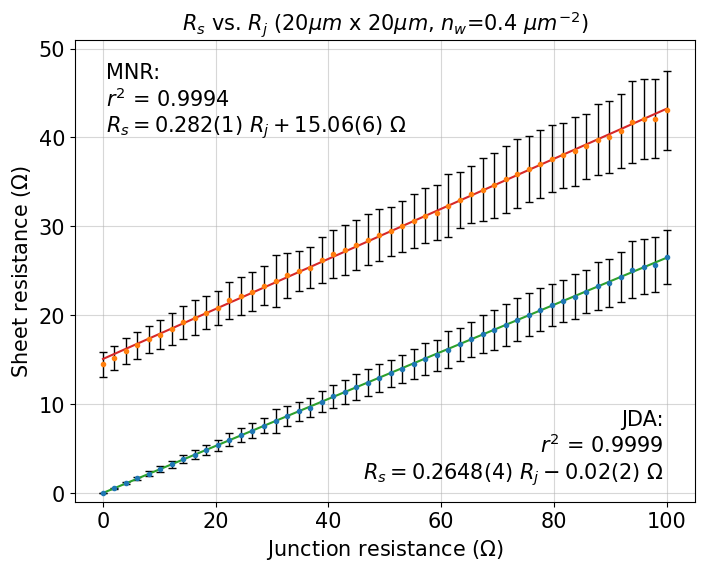}
    \renewcommand{\thefigure}{S1}
    \caption{NWN sheet resistance ($R_s$) as a function of junction resistance ($R_j$). 1000 NWNs were simulated for each of the 50 $(R_j,R_s)$ points obtained using JDA and MNR schemes. NWN dimensions were $20\times \SI{20}{\micro m}$ with a nanowire density of $\SI{0.4}{\micro m^{-2}}$. Each nanowire was set with \SI{7}{\mu m} in length. MNR requires further specifications to account for the inner wire resistances; the NWNs were simulated as made of Ag material, and diameters fixed at \SI{50}{nm}. A \SI{10}{V} voltage difference was applied across \SI{20}{\micro m} long electrodes on the left and right sides. Uncertainties are one standard deviation of the sheet resistances simulated at the given junction resistance point. Best linear fits using model \eqref{eq:SI} were conducted for both trends, and the results are displayed on the panel, including their respective coefficient of determination ($r^2$) and parametric uncertainties.}
    \label{fig:R_s-vs-R_j}
\end{figure}

A well-known result we obtained in \cite{SIrocha2015ultimate} was the linear dependency of the NWN sheet resistance ($R_s$) as a function of the junction resistance ($R_j$), considering static transport regime and a NWN with a fixed wire density, $n_w$. We reproduced this trend using \texttt{MemNetSim} in \autoref{fig:R_s-vs-R_j}. The expected linear trends were obtained for both network nodal representations, JDA and MNR. A two-terminal setup is simulated for $n_w=$ $\SI{0.4}{\micro m^{-2}}$ and, for the sake of simplicity, all junction resistances were set to the same $R_j$ value. Distinct nanowire configurational spatial arrangements can be sorted for the same fixed $n_w$, rendering the need for ensemble analysis for statistically meaningful results. A total of \SI{50000}{} NWNs was simulated, which consisted of 1000 NWN ensembles for each of the 50 linearly spaced junction resistances between \SI{0.001}{\ohm} and \SI{100}{\ohm}. A NWN surface area of $\SI{20}{\micro m} \times \SI{20}{\micro m}$ was generated to randomly disperse the nanowires, and each nanowire was set to be $\SI{7}{\micro m}$ in length. A $\SI{10}{V}$ voltage difference was applied across the network, and MNA was used to find the nodal voltages and current draw with a characteristic resistance of $\SI{10}{\ohm}$ to scale the results. JDA representation does not account for the nanowire inner resistances, therefore, nanowire diameter and segment lengths are not relevant. For MNR, though, the same simulation settings as described above were in place, with the addition of the required resistivity parameter fixed at $\SI{22.6}{n\ohm m}$ and nanowire diameter information fixed at \SI{50}{nm} to account for Ag nanowire inner resistances. Over the numerical data points, we conducted a linear fit to extract the slope and intercept of the trends for both JDA and MNR. The line model is mathematically written as 
\begin{equation}
R_s = a R_j + b \tag{S1} \label{eq:SI}
\end{equation}
\noindent with $a$ being the floating parameter for the slope and $b$ for the intercept. Visually, one can infer that both trends exhibit approximately the same slope, with the difference that the MNR trend is offset by a non-zero intercept. For completeness and comparison, the best-fit line parameters obtained for both representations are indicated in the figure panel, as well as their respective coefficient of determination ($r^2$) and their associated uncertainties as standard errors. For JDA, it is expected that $R_s\rightarrow 0$ as $R_j\rightarrow 0$, but note that the circuit model cannot have $R_j$ exactly at zero. As the junction resistance approaches zero, so does the sheet resistance within the margin of error. This behaviour is expected as the junction resistances are the only source of resistance within this approximation. On the other hand, MNR also carries the nanowire inner resistances, shifting the $R_s$ trend rigidly upwards in resistance value, captured by $b\neq 0$. For the NWN with $n_w=$ $\SI{0.4}{\micro m^{-2}}$, the combined inner resistance contribution from all nanowire segments (excluding dead-ends) resulted in an offset of $b=$ \SI{15.06\pm0.06}{\ohm}, which is also expected from our findings in \cite{SIrocha2015ultimate}. Note that the uncertainties of the data points are one standard deviation of the sheet resistances simulated at that point. One can observe that these uncertainties tend to decrease as the junction resistance decreases. Since the sheet resistance would be sensitive to a single low resistance pathway between the electrodes, this uncertainty would indeed decrease if the simulated distribution of NWNs only contained low junction resistances. It is also worth mentioning that although visually the slopes of the two trends look similar, the quantitative fitting indicated a difference in the best-fit values between JDA and MNR, with $a_{MNR}$ slightly larger than $a_{JDA}$. This suggests that the added resistances of the nanowire themselves cause a slightly faster growth in $R_s$ than without this consideration, and that the combined effects from $R_j$ and $R_{in}$ are weakly linked.

\begin{figure}[H]
    \centering
    \includegraphics[width=0.8\linewidth]{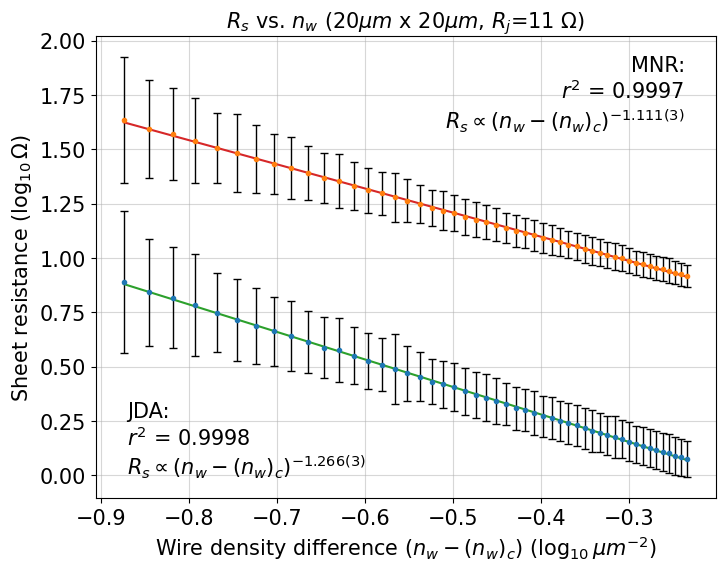}
    \renewcommand{\thefigure}{S2}
    \caption{Log-log plot of NWN sheet resistance ($R_s$) as a function of nanowire density ($n_w$) and critical nanowire density ($(n_w)_c$) difference. 1000 NWNs were simulated for each of the 50 $(n_w,R_s)$ points obtained using JDA and MNR schemes. NWN dimensions were $20\times \SI{20}{\micro m}$ with junction resistances for all junctions fixed at $R_j=$ $\SI{11}{\ohm}$. Each nanowire was set with \SI{7}{\mu m} in length. MNR requires further specifications to account for the inner wire resistances; the NWNs were simulated as made of Ag material, and diameters fixed at \SI{50}{nm}. A \SI{30}{V} voltage difference was applied across \SI{20}{\micro m} long electrodes on the left and right sides. Uncertainties are one standard deviation of the sheet resistances simulated at the given junction resistance point. The critical nanowire density $(n_w)_c$ is the nanowire density in which 50\% of the generated NWNs percolate. A value of $\SI{0.1163}{\micro m^{-2}}$ was used here \cite{SIocallaghan2018transport}, taken from \eqref{eq:constraint} from Li \& Zhang \cite{SIli2009finite}. Uncertainties are one standard deviation of the sheet resistances simulated at the given nanowire density difference point. Best fits using model \eqref{eq:power-law} were conducted for both JDA and MNR curves, and the results are displayed on the panel, including their respective coefficient of determination ($r^2$) and parametric uncertainties.}
    \label{fig:R_s-vs-n_w}
\end{figure}

Another important benchmark is to verify the dependence of the sheet resistance on the nanowire density $n_w$, while holding the junction resistance constant. In a similar manner, \SI{50000}{} NWNs were simulated, which consisted of 1000 samples for each of 50 linearly spaced nanowire densities between \SI{0.25}{\micro m^{-2}} and \SI{0.7}{\micro m^{-2}}. The NWNs generated were $\SI{20}{\micro m} \times \SI{20}{\micro m}$ with a nanowire length of \SI{7}{\micro m}, a junction resistance of \SI{11}{\ohm}. The same \SI{10}{V} voltage difference across the electrodes was applied. Results were scaled with the same characteristic resistance of $\SI{10}{\ohm}$. In the event that a generated NWN for a given nanowire density did not percolate the electrodes, which was more likely for sparser densities, that NWN was omitted from the analysis. JDA and MNR results are illustrated as a log-log plot in \autoref{fig:R_s-vs-n_w}. The sheet resistance as a function of nanowire density is realized as a power law relationship, which matches the relationship previously studied for two-dimensional stick networks \cite{SIli2010conductivity,SIsahimi1983critical}, which identified the relationship
\begin{align}
    R_S \propto (n_w - (n_w)_c)^{-\alpha}
    \tag{S2} \label{eq:power-law}
\end{align}
where $(n_w)_c$ is the critical nanowire density and $\alpha$ is the critical exponent. The critical nanowire density is defined as the nanowire density at which 50\% of the simulated NWNs percolate. That is, 50\% of simulated NWNs have a conducting pathway between the two electrodes. The value used in \autoref{fig:R_s-vs-n_w} was obtained following Li \& Zhang \cite{SIli2009finite} relation
\begin{align}
    (n_w)_c L^2 = \SI{5.63723\pm0.00002}{}
    \tag{S3} \label{eq:constraint}
\end{align}
where $L$ is the nanowire length. For the NWNs simulated with all nanowire lengths equal to $L=$ $\SI{7}{\micro m}$, the critical wire density is thus estimated at $(n_w)_c=$ $\SI{0.1150455\pm0.0000004}{\micro m^{-2}}$. 

Through fitting \autoref{eq:power-law} onto the data points of \autoref{fig:R_s-vs-n_w}, the critical exponent $\alpha$ was found to be \SI{1.266\pm0.003}{} for the JDA NWN type and \SI{1.111\pm0.003}{} for the MNR NWN type. While these are close to literature values of two-dimensional conductive stick networks \cite{SIli2010conductivity,SIsahimi1983critical,SIbalberg1983critical}, the difference between the JDA and MNR critical exponents highlights the effect internal nanowire resistances have in the simulation of NWNs. This result is the expected trend for how $R_s$ should behave with increasing nanowire densities, supporting once more the physical law implementations done in the package.

\end{document}